\documentclass{aa}  

\usepackage{graphicx}
\usepackage{array}
\usepackage{tabularx}
\usepackage{txfonts}
\usepackage[colorlinks=true,     linkcolor=blue, citecolor=blue, filecolor=blue, urlcolor=blue]{hyperref}

\def\ltsima{$\; \buildrel < \over \sim \;$}
\def\simlt{\lower.5ex\hbox{\ltsima}}
\def\gtsima{$\; \buildrel > \over \sim \;$}
\def\simgt{\lower.5ex\hbox{\gtsima}}

\def\msun{{\rm M_{\odot}}}

\def\be{\begin{equation}}
\def\ee{\end{equation}}

\def\del#1{{}}
\def\ltsima{$\; \buildrel < \over \sim \;$}
\def\simlt{\lower.5ex\hbox{\ltsima}}
\def\gtsima{$\; \buildrel > \over \sim \;$}
\def\simgt{\lower.5ex\hbox{\gtsima}}

\usepackage{xcolor}

\newcolumntype{L}{>{$}l<{$}}
\newcolumntype{C}{>{$}c<{$}}
\newcolumntype{R}{>{$}r<{$}}

\usepackage{booktabs}
\usepackage{siunitx}

\begin{document}

   \title{The complex effect of gas cooling and turbulence on AGN-driven outflow properties}
   \titlerunning{AGN outflow coupling}

   \author{K. Zubovas\orcid{0000-0002-9656-6281}\inst{1}, M. Tartėnas\orcid{0009-0006-7373-180X}\inst{1} \and M. A. Bourne\orcid{0000-0003-3189-1638}\inst{2,3}}

   \institute{Center for Physical Sciences and Technology, Saulėtekio al. 3, Vilnius LT-10257, Lithuania\\
              \email{kastytis.zubovas@ftmc.lt}
            \and
             Institute of Astronomy, University of Cambridge, Madingley Road, Cambridge, CB3 0HA, UK 
             \and
            Kavli Institute for Cosmology, University of Cambridge, Madingley Road, Cambridge, CB3 0HA, UK\\ 
             }

   \date{Received ...; accepted ...}

 
  \abstract
   {Accretion onto supermassive black holes at close to the Eddington rate is expected to drive powerful winds, which have the potential to majorly influence the properties of the host galaxy. Theoretical models of such winds can simultaneously explain observational correlations between supermassive black holes and their host galaxies, such as the $M-\sigma$ relation, and the powerful multi-phase outflows that are observed in a number of active galaxies. Analytic models developed to understand these processes usually assume simple galaxy properties, namely spherical symmetry and a smooth gas distribution with an adiabatic equation of state. However, the interstellar medium in real galaxies is clumpy and cooling is important, complicating the analysis.}
   {We wish to determine how gas turbulence, uneven density distribution, and cooling influence the development of active galactic nucleus (AGN) wind-driven outflows and their global properties on kiloparsec scales.}
   {We calculated a suite of idealised hydrodynamical simulations of AGN outflows designed to isolate the effects of turbulence and cooling, both separately and in combination. All simulations initially consisted of a 1 kpc gas shell with an AGN in the centre. We measured the main outflow parameters -- the velocity, the mass outflow rate ($\dot{M}_{\rm out}$), and the momentum ($\dot{p}_{\rm out}c/L_{\rm AGN}$) and energy ($\dot{E}_{\rm out}/L_{\rm AGN}$) loading factors -- as the system evolves over 1.2 Myr and estimated plausible observationally derived values.}
   {We find that adiabatic simulations approximately reproduce the analytical estimates of outflow properties independently of the presence or absence of turbulence and clumpiness. Cooling, on the other hand, has a significant effect, reducing the outflow energy rate by one to two orders of magnitude in the smooth simulations and by up to one order of magnitude in the turbulent ones. The interplay between cooling and turbulence depends on AGN luminosity: in Eddington-limited AGN, turbulence enhances the coupling between the AGN wind and the gas, while in lower-luminosity simulations, the opposite is true. This mainly occurs because dense gas clumps are resilient to low-luminosity AGN feedback but get driven away by high-luminosity AGN feedback. The overall properties of multi-phase outflowing gas in our simulations qualitatively agree with observations of multi-phase outflows, although there are some quantitative differences. We also find that using `observable' outflow properties leads to their parameters being underestimated by a factor of a few compared with real values.}
   {We conclude that the AGN wind-driven outflow model is capable of reproducing realistic outflow properties in close-to-realistic galaxy setups and that the $M-\sigma$ relation can be established without efficient cooling of the shocked AGN wind. Furthermore, we suggest ways to improve large-scale numerical simulations by accounting for the effects of AGN wind.}


   \maketitle
%

\section{Introduction}

Observed supermassive black hole (SMBH) scaling relations such as the $M_{\rm BH}-\sigma$ \citep{Ferrarese00, GultekinEtal09, McConnellMa13, KormendyHo13, Bennert2021ApJ} and $M_{\rm BH}-M_{\rm b}$ \citep{Haering04, McConnellMa13, KormendyHo13} relations provide indirect evidence that feedback from active galactic nuclei (AGN) has an impact on the evolution of the host galaxy.  More direct observational evidence comes in the form of wide-angle cold gas outflows that extend up to $\sim 10$~kpc from the nucleus \citep{SpenceEtAl16}, with mass outflow rates of up to $\sim 1000$ M$_{\odot}$ yr$^{-1}$, momentum flow rates of up to $\sim 50 L_{\rm AGN}/c$, and energy flow rates of $\sim 0.05-10\% L_{\rm AGN}$ \citep{Feruglio2010AA, Rupke2011ApJ,Sturm2011ApJ, Maiolino2012MNRAS, Cicone2014AA, CiconeEtAl2015, Tombesi2015Natur, Gonzalez2017ApJ, Fiore2017AA, Fluetsch2019MNRAS, Lutz2020AA}.

In addition to the large-scale outflows, many active galaxies show evidence of fast nuclear winds, with velocities of $\sim 0.1$~c \citep{PoundsEtal03a,PoundsEtal03b,TombesiEtAl10, Tombesi2010ApJ, Tombesi2013MNRAS}, from blueshifted lines observed in the X-ray band, which have subsequently been dubbed ultra-fast outflows (UFOs). Their kinetic power is comparable to that of the large-scale outflows, suggesting a possible physical connection between the two.

There are several theoretical models that attempt to explain the presence of outflows, such as jet-driven cocoons \citep{Gaibler2012MNRAS, BourneYang2023, TalbotEtAl22, Talbot2024MNRAS}, AGN radiation pressure effects on dusty gas \citep{IshibashiFabian15, ThompsonEtAl15, Costa2018MNRASb}, and AGN wind-driven outflows \citep{costa14, King2015ARAA}. This last framework provides a connection between UFOs and large-scale outflows by showing that UFOs arise as winds from the AGN accretion disc \citep{King2010MNRASb} that subsequently shock against the interstellar medium (ISM) and reach temperatures of $\sim 10^{10}$ K \citep[e.g.][]{King2003ApJ, Zubovas2012ApJ, Faucher2012MNRASb}. The subsequent evolution depends strongly on whether the shocked wind bubble cools efficiently. If it does, then only the wind momentum is transferred to the surrounding gas \citep[e.g.][]{King2003ApJ, King05, Nayakshin2014MNRAS}. The condition for the wind to be able to push away most of the surrounding gas is then a SMBH mass $M_{\rm BH} \propto \sigma^4$, which is close to the observed relation \citep{King2010MNRASa, Zubovas2012MNRASb}. Alternatively, if most of the shocked wind energy is available to drive the ISM, a fast, massive, large-scale outflow develops with properties very similar to those observed \citep{Zubovas2012ApJ, Faucher2012MNRASb}.

In several galaxies, UFOs and large-scale outflows have been discovered simultaneously \citep[e.g.][]{Feruglio2015AA, Tombesi2015Natur, Veilleux2017ApJ}. Although in some cases the energy rates of the two components match almost exactly \citep[e.g. in Mrk 231; see][]{Feruglio2010AA, Feruglio2015AA}, typically there is a significant discrepancy between the two values \citep{Marasco2020AA}. There are two primary explanations for this. The first involves AGN luminosity variations over the dynamical timescale of the outflow, $t_{\rm flow} \sim R/v_{\rm out} \sim 1 R_{\rm kpc} v_8^{-1}$~Myr, where $v_8 \equiv v_{\rm out} / 1000$~km~s$^{-1}$. Since outflow properties cannot change on timescales much shorter than dynamical, it follows that they correlate better with the average AGN luminosity over that time \citep{Zubovas2020MNRAS}. Individual AGN episodes should not last longer than a few times $10^5$~yr \citep{King2015MNRAS, Schawinski2015MNRAS}, so the instantaneous luminosity we observe is generally not representative of the long-term average. UFO properties, on the other hand, change on much shorter timescales, of years or less \citep{King2015ARAA}, and so trace the AGN luminosity much more closely. The inclusion of this variability allows models to reproduce the scatter of large-scale outflow properties when compared against present-day AGN luminosity \citep{Zubovas2020MNRAS, Zubovas2022MNRAS}.

The second possible explanation of the variation of outflow properties with regard to UFO and/or AGN properties is uneven coupling in different galaxies. Generally, the ISM exhibits clumping on various scales, with cold dense clouds and filaments embedded in a more diffuse hot component. In this case, most of the shocked wind energy can escape via paths of least resistance \citep{wagner13, Nayakshin2014MNRAS, Zubovas2014MNRASb, bourne14,BieriEtAl17a}, namely the low-density channels in the ISM, leaving the cold gas in its wake exposed mostly to the wind momentum. In fact, the fraction of wind energy affecting different clumps runs the gamut from zero to unity \citep{Zubovas2014MNRASb}. Many clumps can be compressed by the passage of the outflow, enhancing star formation \citep[][submitted]{LauzikasZubovas2024}, facilitating its infall and reducing the effective cross-section of dense clumps exposed to the wind \citep{bourne14}. This scenario also helps explain a particular shortcoming of the more idealised, spherically symmetric, wind-driven outflow model. Within that formulation, a momentum-driven outflow can only form if the shocked wind cools efficiently, mainly via the inverse Compton effect \citep{King2003ApJ}. The transition between efficient and inefficient cooling should occur on scales of several hundred parsecs; however, the expected signature of radiative cooling on these scales is not observed \citep{Bourne13}. In addition, when one considers that the shocked wind plasma is likely two-temperature, with protons carrying most of the energy, it becomes clear that inverse Compton cooling is very inefficient \citep{Faucher2012MNRASb}; as such, energy-driven outflows should form very close to the SMBH, clearing gas from the galaxy and preventing the SMBH from growing. The possibility of both momentum- and energy-driven outflows existing and  affecting different phases of the gas resolves this issue.

While the behaviour of dense gas in multi-phase outflows has received some attention, the large-scale properties of outflows launched in multi-phase media remain relatively unexplored. In particular, it is uncertain whether the global properties -- the kinetic powers, momentum, and mass outflow rates -- of the large-scale outflows produced in such systems agree with observational results. Intuitively, we would expect the dense material that comprises the majority of the outflowing mass to move with lower velocities than predicted in the spherically symmetric scenario, with a corresponding decrease in momentum and energy rates and corresponding loading factors. This picture is complicated by the generally non-linear heating and cooling processes, which lead to the outflowing gas changing phase as the outflow evolves \citep{Zubovas2014MNRASa, Richings2018MNRAS, Richings2018MNRASb}. Although early detections of massive outflows \citep{Feruglio2010AA, Sturm2011ApJ, Rupke2011ApJ, Maiolino2012MNRAS} generally agreed quite well with the spherically symmetric prediction \citep{Cicone2014AA}, recent larger outflow samples show much larger scatter around these predictions, primarily towards lower values for a given AGN luminosity \citep{Fluetsch2019MNRAS, Lutz2020AA}.

In this study we used numerical simulations of idealised systems to quantify the distribution and evolution of three major outflow parameters -- the mass outflow rate, momentum loading factor, and energy loading factor -- in both smooth and turbulent ISM, with and without gas cooling. We show that turbulence by itself does not impact the global parameters (i.e. the energy injected by the AGN couples equally well to smooth and clumpy gas distributions). Cooling, on the other hand, has a profound effect even though we assume that the shocked AGN wind remains adiabatic. The effect of cooling is intertwined with that of gas clumpiness and shock heating by the outflow: in smooth systems, cooling reduces the outflow energy rate by one to two orders of magnitude, with a higher AGN luminosity leading to less efficient cooling; in turbulent systems, the effect is mitigated since some gas can be heated to very high temperatures, leading to inefficient cooling. In general, the mass outflow rates and the momentum and energy loading factors we obtain in simulations with both turbulence and cooling agree quite well with corresponding values of real outflows, and the differences in the host gas density structure go a long way in explaining the observed variety of loading factors.

We briefly review the analytical derivation of outflow parameters in Sect. \ref{sec:analytical} and present the simulation setup in Sect. \ref{sec:sims} and the results in Sect. \ref{sec:results}. We discuss the implications of our results in Sect. \ref{sec:discuss}, focusing on their applicability to interpreting outflow observations, the implementation of AGN feedback in numerical simulations, and the shortcomings and possible improvements of our simulations. We summarise the main results and conclude in Sect. \ref{sec:sum}.

\section{Analytical arguments and models}
\label{sec:analytical}

The expansion of energy-driven outflows in a homogeneous medium has been detailed in several papers, including reviews by \cite{King2015ARAA} and \cite{Zubovas2019GReGr}; here we briefly summarise the main results before considering how we would expect the results to change in an inhomogeneous, clumpy gas distribution. First, it is assumed that the SMBH has reached its formal $M_{\sigma}$ mass \citep{King05},
\begin{equation}
M_{\sigma}=\frac{f_{\rm c}\kappa}{\pi G^{2}}\sigma^{4}=3.7\times 10^{8}\sigma_{200}^{4}{\rm M}_{\odot},
\label{eq:m_sig}
\end{equation}
where $f_{\rm c}=0.16$ is the cosmological baryon fraction, $\kappa = 0.4$~cm$^2$~g$^{-1}$ is the electron scattering opacity and $\sigma_{200}$ is the bulge velocity dispersion in units of $200$ km s$^{-1}$. Secondly, feedback is Eddington-limited and the wind has momentum and kinetic energy rates of
\begin{equation}
\dot{p}_{\rm w} = \frac{L_{\rm Edd}}{c}
\end{equation}
{and} 
\begin{equation}
\dot{E}_{\rm w} = 0.05L_{\rm Edd}.
\end{equation}

In the regime in which the wind shock is unable to cool efficiently (i.e. the energy-driven regime), \citet{Zubovas2012ApJ} show that most of the wind energy is transferred to the kinetic energy of the outflow such that
\begin{equation}
\frac{1}{2}\dot{M}_{\rm out}v_{\rm out}^{2}\simeq\frac{1}{2}\dot{M}_{\rm w}v_{\rm w}^{2}
\label{eq:mdot_full}
,\end{equation}
and hence
\begin{equation}
  \dot{p}_{\rm out} = \dot{p}_{\rm w}\left(\frac{\dot{M}_{\rm out}}{\dot{M}_{\rm w}}\right)^{1/2} = \frac{L_{\rm Edd}}{c}f_{\rm L}^{1/2},
  \label{eq:pdot_full}
\end{equation}
where the AGN wind mass loading ratio 
\begin{equation}
f_{\rm L}=\frac{\dot{M}_{\rm out}}{\dot{M}_{\rm w}}=\left(\frac{2\epsilon_{\rm r} c}{3\sigma}\right)^{4/3}\left(\frac{f_{\rm g}}{f_{\rm c}}\right)^{2/3}\frac{l^{1/3}}{\dot{m}}.
\label{eq:mass_loading_full}
\end{equation}
In this equation, $f_{\rm g}$ and $f_{\rm c}$ are the gas fraction and cosmological baryon fraction, respectively, $\epsilon_{\rm r}$ is the radiative efficiency of accretion, $l=L_{\rm AGN}/L_{\rm Edd}$ and $\dot{m}=\dot{M}_{\rm w}/\dot{M}_{\rm Edd}$. For typical galaxy and wind parameters, assuming $f_{\rm g}=f_{\rm c}$, $f_{\rm L}\sim400$, and so momentum boosts of $\sim20$ and mass outflow rates of several times $10^2-10^3 \msun$ yr$^{-1}$ can be reached.

An important point to note is that the model of \citet{Zubovas2012ApJ} assumes that all of the gas is swept up into the outflow. However, let us now assume that the ISM can be split into hot diffuse and cold dense components. We define the parameter
\begin{equation}
\eta_{\rm hot}=\frac{f_{\rm g}-f_{\rm cold}}{f_{\rm g}}
\end{equation}
as the fraction of gas in the hot phase. Furthermore, if we assume that only the hot phase can be easily swept up in the energy-driven outflow, with the cold high-density clumps left largely intact \citep[e.g.][]{bourne14}, we can define the hot gas mass loading factor as
\begin{equation}
f_{\rm L, hot}= \left(\frac{f_{\rm g, hot}}{f_{\rm g}}\right)^{2/3} f_{\rm L} = \eta_{\rm hot}^{2/3}f_{\rm L}.
\end{equation}
Replacing $f_{\rm L}$ with $f_{\rm L, hot}$ in the above analysis provides an analytical expectation that when the ISM is clumpy, with large fractions of the gas content contained in dense cold clumps, mass outflow rates and momentum boosts should be considerably tempered when compared with spherically symmetric calculations that assume a homogeneous gas distribution. It is important to note that there is, however, a degeneracy between $\eta_{\rm hot}$ and $f_{\rm g}$ and it is also likely that $\eta_{\rm hot}$ is a (possibly complicated) function of $f_{\rm g}$, with gas-rich systems likely being clumpier and hence having smaller values of $\eta_{\rm hot}$. Furthermore, as pointed out in \citet{Nayakshin2014MNRAS}, from the cooling function of \citet{Sutherland93}, the hot gas fraction can be given as $\eta_{\rm hot}\sim 2.5\times 10^{-3}\sigma_{200}^{5/2}R_{\rm kpc}Z^{-0.6}$, where $Z$ is metallicity in solar units and $R_{\rm kpc}$ is the radial position within the host galaxy. As a result, more massive, larger, and/or less metal-rich galaxies should have larger hot gas fractions. This also suggests that as outflows move to larger radii, the mass loading and momentum boosts should increase. However, the reality is likely different from the simple picture painted here; this analysis assumes an isothermal gas density distribution, which is only an approximation to the true gas distribution in galaxies. On top of this, the outflow itself will impact $\eta_{\rm hot}$ as it propagates, through a combination of ablating and disrupting dense gas clouds and/or compressing gas to high densities as it is swept up. As such, we expect $f_{\rm L}$ to be a complicated function of feedback and galaxy properties.

\section{Numerical simulations} 
\label{sec:sims}

\begin{table*}
  \centering
  \def\arraystretch{1.2}
  \caption{Summary of simulations, including key results.}
  \begin{tabular}{@{}lccccccc@{}}    
    \toprule
    Run & gas distribution & cooling? & $L_{\rm AGN}  $ & $v_{\rm out} $ & $\dot{M}_{\rm out}$ & $\dot{p}_{\rm out} c/L_{\rm AGN}$ & $\dot{E}_{\rm k, out}/L_{\rm AGN}$ \\  
    & & & $\left(\text{erg s}^{-1}\right)$ & $\left(\text{km s}^{-1}\right)$ & $\left(\msun \text{ yr}^{-1}\right)$ & & $\left(\times 10^{-3}\right)$ \\
    \midrule
    SmthAdiaL0.3 & smooth & no & $3.78 \times 10^{45}$ &  350 & 642 (348) & 11.3 (6.1) & 7.64 (4.14) \\
    SmthAdiaL1.0 & smooth & no & $1.26    \times 10^{46}$ & 582 & 966 (671) & 8.44 (5.86) & 9.34 (6.49) \\
    SmthCoolL0.3 & smooth & yes & $3.78 \times 10^{45}$ & 96.2 & 229 (166) & 1.1 (0.8) & 0.20 (0.15) \\
    SmthCoolL1.0 & smooth & yes & $1.26 \times 10^{46}$& 304 & 485 (402) & 2.21 (1.83) & 1.23 (1.02) \\
    TurbAdiaL0.3 & turbulent & no & $3.78 \times 10^{45}$ & 312 & 641 (313) & 10 (4.89) & 8.70 (4.26) \\
    TurbAdiaL1.0 & turbulent & no & $1.26 \times 10^{46}$ &  556 & 1030 (630) & 8.64 (5.26) & 11.22 (6.83) \\
    TurbCoolL0.3 & turbulent & yes & $3.78 \times 10^{45}$ &  140 & 116 (23.3) & 0.817 (0.164) & 0.78 (0.16) \\
    TurbCoolL1.0 & turbulent & yes & $1.26 \times 10^{46}$ & 297 & 710 (309) & 3.18 (1.38) & 3.38 (1.47) \\
    \hline
    analytical-L0.3 & - & - & $3.78 \times 10^{45}$ & $590 - 790$ & $565-760$ & $17-30$ & $0.016-0.04$\\
    analytical-L1.0 & - & - & $1.26 \times 10^{46}$ & $885-1180$ & $850-1130$ & $11-20$ & $0.017-0.039$ \\
    \hline
    \vspace{0.1cm}
  \end{tabular}
  \vspace{0.3cm}
  \begin{minipage}{\textwidth}
    \small
    \textbf{Notes:} The first column shows the simulation name, the second - the type of gas distribution used, the third - the presence or absence of radiative cooling, and the fourth shows the AGN luminosity. The subsequent columns show the results: mass-weighted outflow velocity, time-averaged mass outflow rate, and momentum and energy loading factors. In the final three columns, the first value is evaluated assuming $\tau=\tau_{\rm AGN}$, while the one in parentheses assumes $\tau = \tau_{r/v}$. All results evaluated at $t = 0.5$~Myr (see the main text for the reasoning). Two rows at the bottom show the analytical predictions for the main results, based on \citet{Zubovas2012ApJ}. The lower velocity is that of the contact discontinuity, while the higher corresponds to the outer shock; other parameter ranges correspond to these two velocities.
    
  \end{minipage}
  \label{tab:sims}
\end{table*}

While analytical models provide physical intuition, they can take us only so far and as such we now turn to numerical simulations. We adopted a similar setup and methodology to that of simulations previously presented in \citet{Bourne2015MNRAS}, \citet{ZBN2016}, and \cite{Zubovas2023MNRAS}. A modified version of the hydrodynamical code {\sc GADGET-3} \citep{Springel05}, which incorporates the `SPHS' \citep{Read10, ReadHayfield12}  flavour of SPH (smoothed particle hydrodynamics + switch) and the Wendland kernel \citep{Wendland95,DehnenAly12}, was used for the simulations. The code employs adaptive hydrodynamical smoothing and gravitational softening lengths; the neighbour number is $N_{\rm ngb}=100$. The SPH particle mass is $m_{\rm SPH}=940 \msun$.

\subsection{Initial conditions} \label{sec:ic_gen}

We run two groups of simulations: smooth (labelled `Smth') and turbulent (`Turb').  In the smooth simulations, the initial conditions consist of a shell of gas between $r_{\rm in} = 0.1$~kpc and $r_{\rm out} = 1$~kpc. The density falls with radius as $\rho \propto r^{-2}$ and the total mass is $M_{\rm gas} = 9.4\times 10^{8} \msun$, tracked with $N = 10^{6}$ SPH particles. We neglected the self-gravity of the gas to avoid the spurious fragmentation of dense clumps. The shell is embedded in a static singular isothermal sphere background potential with a one-dimensional velocity dispersion of $\sigma_{\rm b} = 142$ km s$^{-1}$, corresponding to a background mass $M_{\rm b} = 9.4\times10^9 \, \msun$. Finally, a black hole (BH) particle of mass $M_{\rm bh}=M_{\sigma}\left(\sigma_{\rm b}\right)=10^{8}$ M$_{\odot}$ is included at the centre of the simulation domain and kept fixed throughout the simulation. All gas particles that fall closer than $r_{\rm acc} = 0.01$~kpc from the BH particle are removed from the simulation.

In the turbulent simulations, the extent of the gas shell is the same, but it is given a turbulent density distribution. To achieve this, we ran a `development' simulation that was identical to the smooth simulations except for the following differences. The gas shell initially extends between $r_{\rm in, d} = 0.01$~kpc and $r_{\rm out} = 1$~kpc, has a total mass larger by a factor of $\sim 1.4$ and is given a turbulent velocity field. Turbulence is generated using the method of \citet{DubinskiEtAl95} and \citet{HobbsEtal11}, which produces a divergence-free turbulent velocity spectrum. The initial characteristic velocity is $\sigma_{\rm t} = 149$~km~s$^{-1} = \sigma_{\rm b}\left(1+M_{\rm gas}/M_{\rm b}\right)^{1/2}$. After 1 Myr, we stopped the simulation and removed the gas in the central $r_{\rm in} = 0.1$~kpc. The leftover shell now contains $N \simeq 10^6$ particles within $r_{\rm out} = 1$~kpc, which we used as the initial conditions for the turbulent simulations. We repeated this process four times with different random seeds and performed simulations with the four sets of initial conditions. The background potential and BH particle properties in the development and turbulent simulations are identical to those of the smooth simulations.

\subsection{Gas thermodynamics} \label{sec:thermodynamics}

Since we are interested in understanding how outflow properties are affected by a more realistic gas density distribution, we adopted a two-step approach to make the simulation more realistic. The first step is the creation of turbulent initial conditions, as described above. The second step is moving from an adiabatic equation of state, as assumed in the analytical calculations, to one with a realistic radiative cooling prescription, described below. We ran both adiabatic (labelled `Adia') and cooling (labelled `Cool') simulations. In the adiabatic simulations, the initial gas temperature was set to the virial temperature of the background gravitational potential, $T\sim9.3\times10^5\,$K; we verified that using a lower initial temperature makes no difference to our results.

In the simulations with radiative cooling, we used a combination of two sub-resolution prescriptions. Below temperatures of $10^{4}$ K, we adopted the function of \citet{Mashchenko08}, which allows gas to cool to 20 K. This cooling function is designed to incorporate the effects of atomic, molecular and dust-mediated cooling in Solar-metallicity gas. This means cooling is particularly efficient; together with the adiabatic simulations, our setup brackets a large range of `interesting' cooling regimes. Above $10^{4}$ K, we used the prescription of \citet{SOCS2005}, which models the heating and cooling of optically thin gas due to a typical AGN radiation field. We modified this function by neglecting the effect of Compton cooling, a change appropriate for the expected two-temperature plasma nature of the hottest outflowing gas \citep[see][]{Faucher2012MNRASb, Bourne13}. The optically thin approximation is unrealistic for the densest clumps, and so our simulations over-predict the heating rate in these regions and hence under-predict the total mass of cold gas. However, in terms of global outflow properties and time-integrated star formation rates, this has a negligible effect \citep{ZubovasBourne17}.

\subsection{AGN episode and feedback injection} \label{sec:injection}

In all simulations, an AGN is turned on at the start and allowed to continue for the duration of the simulation, until the gas is cleared from the central kiloparsec or falls into the SMBH particle. We ran simulations with different AGN luminosities, labelled `L\#', where `\#' refers to the luminosity in units of the Eddington luminosity $L_{\rm Edd} = 1.26 \times 10^{46}$~erg/s. Our fiducial simulations are `L1' (i.e. $L_{\rm AGN} = L_{\rm Edd}$), but we also checked whether the results depend on luminosity by running `L0.3' simulations (i.e. with $L_{\rm AGN} = 0.3L_{\rm Edd}$). We also ran four control simulations without an AGN. The AGN affects the gas in two ways: by heating (described in Sect. \ref{sec:thermodynamics} above) and by producing feedback in the form of a fast wind, which we tracked using a novel grid-based scheme. The prescription is described in detail in a companion paper (Tart\.{e}nas and Zubovas, in prep.); here we present a summary of the salient points.

The method works by propagating the AGN wind on a static grid, which allows for a quick coupling with co-spatial particles using a spatial hash-like method and a sorted distance matrix. Spatial hashing is used widely outside astronomy \citep[e.g. in collision detection and computer vision;][]{Hastings2005, CollisionDetection, 3Dperception2023} and was suggested as an effective algorithm for friends-of-friends group finding in astrophysical simulations (\citealt{Creasey2018}; however, this has been contested by \citealt{Gadget4_2021}).

We used a modified version of the spherical rectangular equal-area grid \citep{SREAG} and neglected the modification of the $\theta$ coordinate. While this modification results in a variation of cell areas of the order of a few per cent, it allows for a simple and quick determination  of integer grid indices ($r, \theta, \phi$) for SPH particles:
\begin{align}
    r &=\left\lfloor r_{\rm p} / \Delta r\right\rfloor, \\
    \theta &= \left\lfloor\theta_{\rm p} / \Delta \theta\right\rfloor, \\
    \phi &= \left\lfloor\phi_{\rm p} / \Delta \phi(\theta)\right\rfloor,
\end{align}
where $r_{\rm p}, \theta_{\rm p}, \phi_{\rm p}$ are SPH particle's spherical coordinates, $\Delta r, \Delta \theta, \Delta \phi$ are the respective grid step sizes, and $\lfloor x \rfloor$ denotes the floor function.

We used a simple discrete-step approach for wind propagation. First, wind was injected into selected 0-th shell cells, the total energy determined by $L_{\rm AGN}$ and the SMBH timestep. We assumed that AGN wind travels radially outwards at a constant velocity $v_{\rm wind}=0.1\rm{c}$ \citep{King2003ApJ, King2010MNRASa}. Wind propagation was coupled to the SMBH timestep, with timestep limits set by $C \Delta r / v_{\rm wind}$, where $C=0.4$ is a Courant-type factor. 

Feedback is distributed to SPH particles in proportion to those particles' contribution to the overall density field at the centre of the grid cell. Each wind variable - in our case, energy and momentum - is transferred independently. In the case of momentum injection, if the particle position is outside the cell in question, momentum is injected in the direction of the cell rather than the particle.

\subsection{List of simulations}

A summary of the main simulation parameters and salient results is given in Table \ref{tab:sims}. The first column gives the simulation name, and the next three give the main parameters encoded in the name: the type of gas distribution (smooth or turbulent), the presence of radiative cooling (yes or no) and the AGN luminosity. The subsequent columns give the primary results, in the turbulent case averaged over variations in the initial conditions. They are, in order, the outflow velocity, mass outflow rate, momentum loading factor, and energy loading factor. These values are averaged over the extent of the outflow, with the first (second) value in the last three columns using $\tau_{\rm AGN}$ ($\tau_{r/v}$) when averaging (see Eq. \ref{eq:mdot_average} and the text below it for the precise definitions used). All values are evaluated at $t=0.5$~Myr, when the outflow has already developed but has not yet broken out of the initial shell even in the turbulent simulations. The bottom two rows of the table show analytical predictions for the main results, based on equations in \citet{Zubovas2012ApJ}, using either the velocity of the contact discontinuity (lower) or the outer shock (higher) for calculations. We note that the analytically estimated values are upper limits, since the velocity calculation depends on the assumption that $v_{\rm out} \gg \sigma_{\rm b}$, which allows us to ignore one of the terms in the equation of motion \citep{Zubovas2012ApJ} and the other quantities are calculated using this $v_{\rm out}$ estimate.

\section{Results} 
\label{sec:results}

\begin{figure*}
\includegraphics[width=\textwidth, trim={0 10 0 10}, clip]{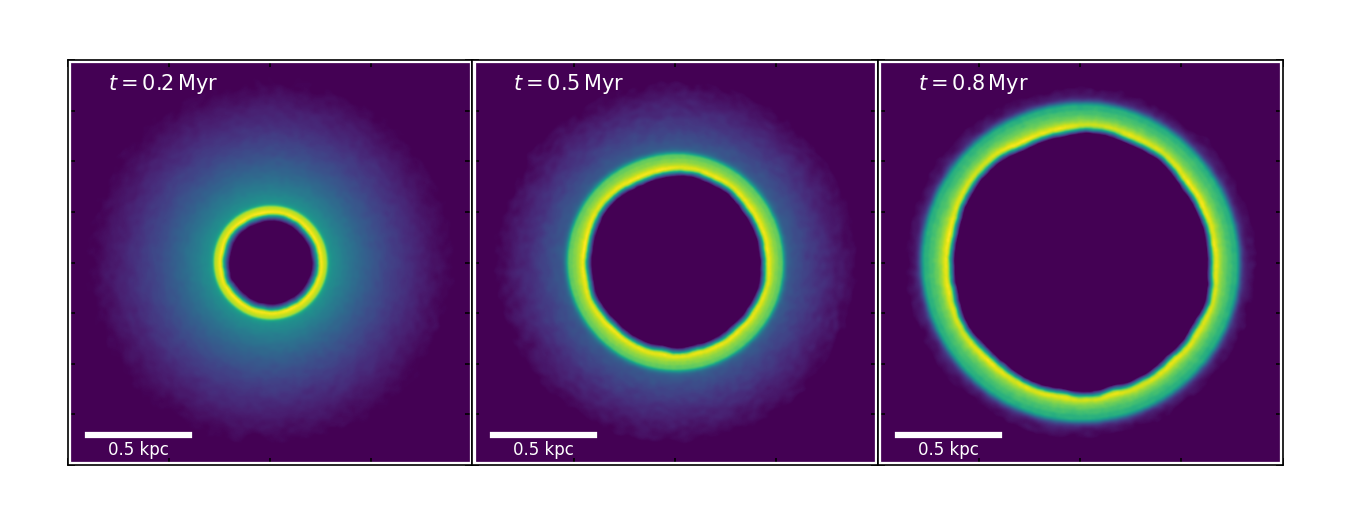}
\caption{Gas density integrated through a $0.1$ kpc thick slice in simulation SmthAdiaL1.0. The three panels show, from left to right, $t = 0.2$~Myr, $t = 0.5$~Myr, and $t = 0.8$~Myr; brighter colours represent higher gas density.}
\label{fig:smooth_density}
\end{figure*}

\begin{figure}
\includegraphics[width=0.48\textwidth, trim={0 33 0 15}, clip]{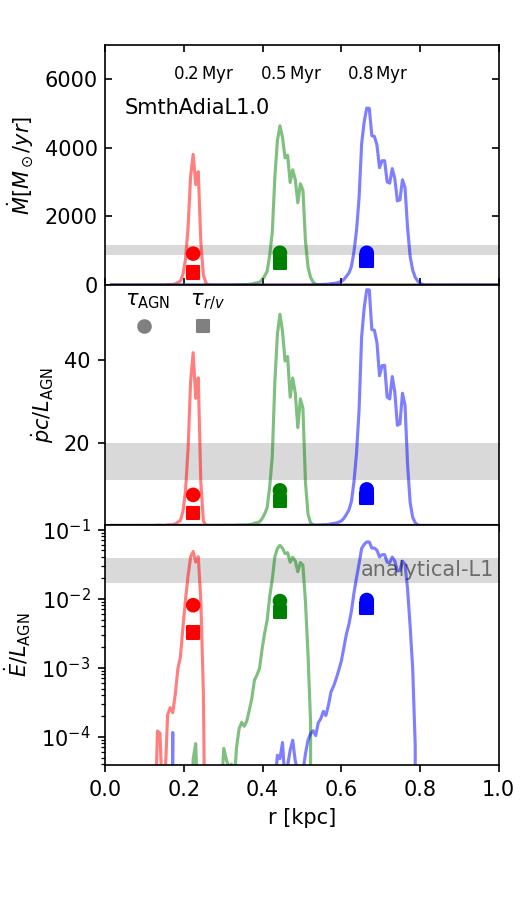} 
\caption{Radial profiles of the mass outflow rate (top), momentum loading factor (middle), and energy loading factor (bottom) in simulations SmthAdiaL1.0 (solid lines and dark points) at $t = 0.2$~Myr (red), $t = 0.5$~Myr (green), and $t = 0.8$~Myr (blue). Lines show the radial profiles calculated using Eqs. \ref{eq:Mdot_radial}, \ref{eq:pdot_radial}, and \ref{eq:Edot_radial}, the latter two converted into loading factors. Points show the globally averaged values calculated using $\tau_{\rm AGN}$ (circles) and $\tau_{r/v}$ (squares). Grey-shaded areas show the range of analytical estimates.}
\label{fig:smooth_profiles}
\end{figure}

We first describe the morphological evolution of gas in all four groups of simulations, as well the evolution of the mass outflow rate ($\dot{M}_{\rm out}$), the momentum loading factor ($\dot{p}_{\rm out} c / L_{\rm AGN}$), and the energy loading factor ($\dot{E}_{\rm out} / L_{\rm AGN}$) with radius and with time. We calculated the radial profiles by summing up relevant particle properties over thin shells:
\begin{equation} \label{eq:Mdot_radial}
\dot{M}_{\rm out}(r)\Delta r=\sum_{i}{m_{i}v_{{\rm rad, }i}},
\end{equation}
\begin{equation} \label{eq:pdot_radial}
\dot{p}_{\rm out}(r)\Delta r=\sum_{i}{m_{i}v_{{\rm rad}, i}^{2}},
\end{equation}
and
\begin{equation} \label{eq:Edot_radial}
\dot{E}_{\rm k, out}(r)\Delta r=\sum_{i}{\frac{1}{2}m_{i}v_{{\rm rad}, i}^{3}},
\end{equation}
where $m_{i}$ and $v_{{\rm rad, }i}$ are the mass and radial velocity of the $i$th particle, and the sum is performed over all outflowing particles in the radial bin between $r$ and $r+\Delta r$; we use $\Delta r = 0.005$~kpc throughout this paper, but the results are insensitive to the precise value as long as $\Delta r \ll 0.1$~kpc. We define a particle to be outflowing if it has $v_{\rm rad} > 10$~km~s$^{-1}$. The chosen slightly higher threshold value excludes particles that are just nominally outflowing due to stochastic scatter in velocity. We also subtracted the corresponding values calculated in the control (L0) simulation to remove the effect of spurious gas motions; in practice, this is only relevant for the turbulent simulations. The total momentum of the outflow can then be calculated as
\begin{equation}
p_{\rm out} = \sum_{r}{\dot{M}_{\rm out}\left(r\right) \Delta r},
\end{equation}
and the total energy as
\begin{equation}
E_{\rm out} = \frac{1}{2}\sum_{r}{\dot{p}_{\rm out}\left(r\right) \Delta r}.
\end{equation}

We also defined the global radially and time-averaged values of outflow properties. The mass flow rate is given by
\begin{equation} \label{eq:mdot_average}
    \dot{M}_{\rm out, ave} = \frac{M_{\rm out, corr}}{\tau_{\rm out}} = \frac{M_{\rm out} - M_{\rm out, L0}}{\tau_{\rm out}},
\end{equation}
where $M_{\rm out, corr}$ is the total outflowing mass corrected for spurious motions by subtracting its value in the corresponding control simulation and $\tau_{\rm out}$ is the age of the outflow.  The latter can be defined in two ways. Firstly, from the simulation we directly know the age of the AGN outflow and can simply set $\tau_{\rm AGN}=t$, where $t$ is the simulation time. However, this information is not directly available to observers, where it is common to estimate the flow time as $\tau_{r/v}=R_{\rm out}/v_{\rm out}$. The exact values used for $R_{\rm out}$ and $v_{\rm out}$ can vary between studies and depend upon what assumptions are made about the outflow properties. Such ideas have already been discussed extensively in the literature \citep[see e.g.][]{RupkeEtAl05,Cicone2014AA,Veilleux2017ApJ}, including the addition of numerical factors to take the outflow geometry into account \citep[e.g.][]{Maiolino2012MNRAS} and the decomposition of the outflow into individual `clumps' \citep[e.g.][]{CiconeEtAl2015, Bischetti2019AA}. For the simple homogeneous case in which all gas is swept into a shell, $R_{\rm out}$ and $v_{\rm out}$ could simply be set to the shell radius and velocity. However, here we have two factors to consider: firstly, in the turbulent simulations we do not have a simple, single-velocity thin-shell outflow, and secondly, our ISM shell has an initial inner cavity with $r_{\rm in}=100$~pc. Therefore, we set $R_{\rm out}=\overline{r}_{\rm out}-r_{\rm in}$ and $v_{\rm out}=\overline{v}_{\rm out}$, which are the mass-weighted average outflow radius (corrected for the shell inner edge) and radial velocity, respectively. Finally, the averaged momentum and energy rates are then calculated as $\dot{p}_{\rm out, ave}=\dot{M}_{\rm out, ave}\overline{v}_{\rm out}$ and $\dot{E}_{\rm out, ave}=0.5\dot{M}_{\rm out, ave}\overline{v^2}_{\rm out}$, respectively, where $\overline{v^2}$ is the mass-weighted average squared radial velocity.

\subsection{Gas morphology and outflow rates}

\subsubsection{Smooth adiabatic simulations} \label{sec:results_S_ad}

\begin{figure}
\includegraphics[width=0.47\textwidth] {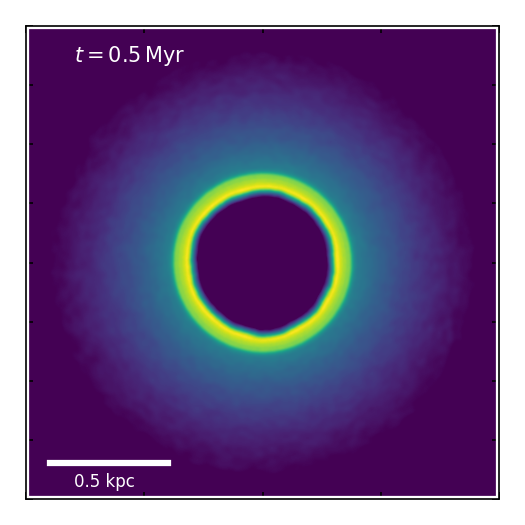} 
\caption{Gas density integrated through a $0.1$ kpc thick slice at $t = 0.5$~Myr in simulation SmthAdiaL0.3.}
\label{fig:smooth_density_L0.3}
\end{figure}


\begin{figure*}
\includegraphics[width=\textwidth, trim={0 10 0 10}, clip]{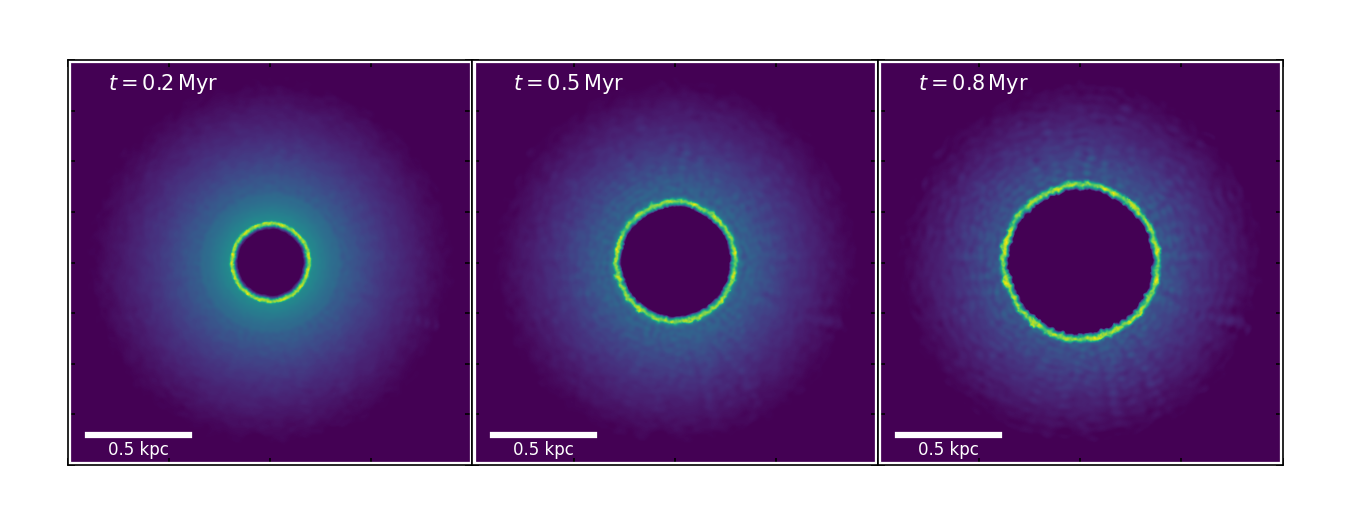}
\caption{Same as Fig. \ref{fig:smooth_density} but for the SmthCoolL1.0 simulation.}
\label{fig:smooth_cooling_density}
\end{figure*}

\begin{figure}
\includegraphics[width=0.47\textwidth]{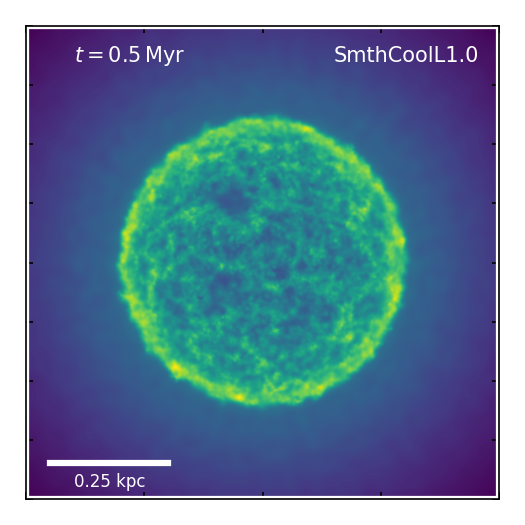} 
\caption{Gas density integrated throughout the simulation volume in simulation SmthCoolL1.0 at $t = 0.5$~Myr.}
\label{fig:smooth_cooling_density_integrated}
\end{figure}

\begin{figure}
\includegraphics[width=0.48\textwidth, trim={0 33 0 15}, clip]{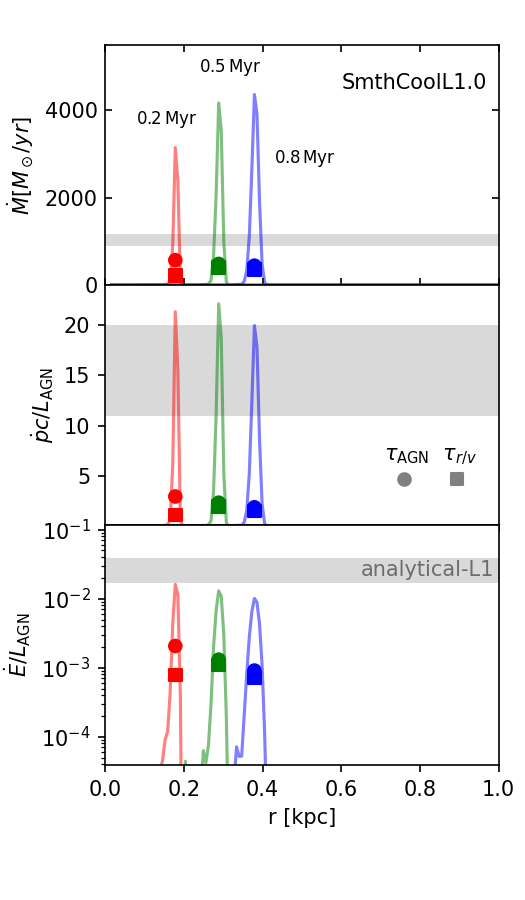} 
\caption{Same as Fig. \ref{fig:smooth_profiles} but for the SmthCoolL1.0 simulation.}
\label{fig:smooth_cooling_profiles}
\end{figure}


\begin{figure*}
\includegraphics[width=\textwidth, trim={0 10 0 10}, clip]{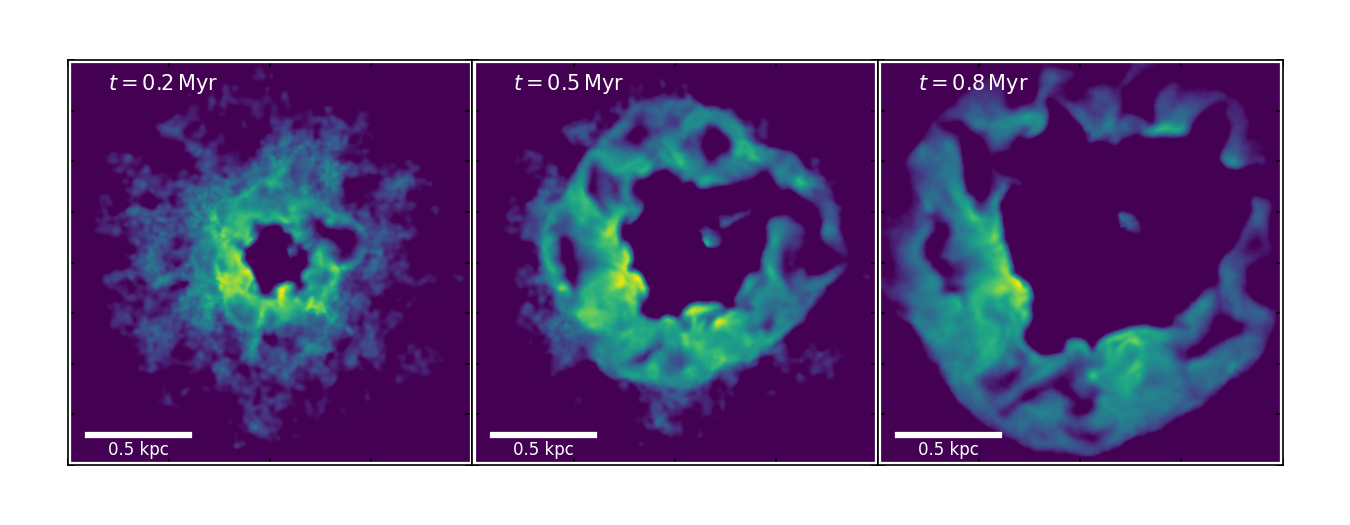}
\caption{Same as Fig. \ref{fig:smooth_density} but for the TurbAdiaL1.0 simulation.}
\label{fig:turbulent_density}
\end{figure*}

\begin{figure}
\includegraphics[width=0.48\textwidth, trim={0 33 0 15}, clip]{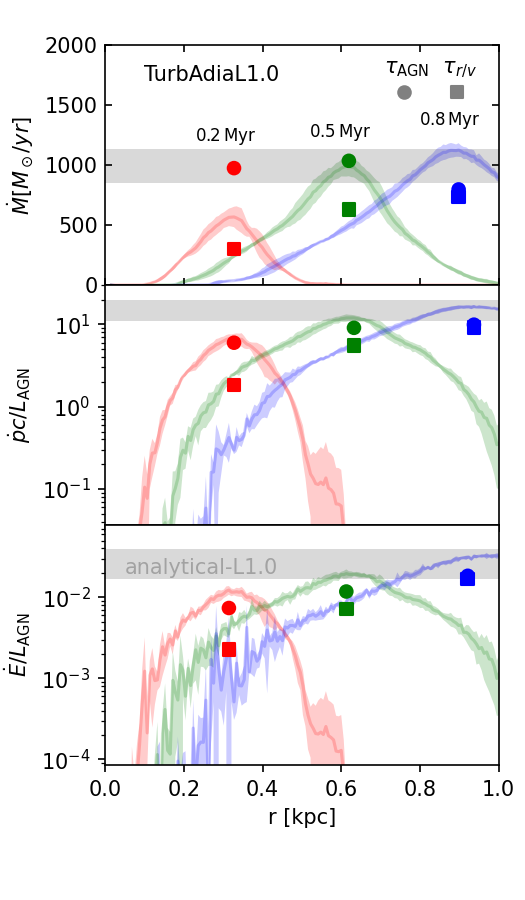} 
\caption{Same as Fig. \ref{fig:smooth_profiles} but for the TurbAdiaL1.0 simulation. Shaded areas show a standard deviation from a sample of four simulations that differ only in terms of the random seed used to generate turbulence.}
\label{fig:turbulent_profiles}
\end{figure}

We show the density distribution of a thin ($0.1$ kpc thickness) slice of the simulation SmthAdiaL1.0 as it evolves in Fig. \ref{fig:smooth_density}, with the left panel showing a snapshot at $t =0.2$~Myr, the middle at $0.5$~Myr and the right at $0.8$~Myr. As expected, the outflow expands spherically symmetrically, with minor variations only due to finite particle number. The peak density, which corresponds approximately to the contact discontinuity, quickly reaches a velocity $v_{\rm c.d.} \simeq 700$~km~s$^{-1}$, while the outer edge expands with a velocity $v_{\rm out} \simeq 900$~km~s$^{-1}$. These values are lower than the analytical estimate by $\sim 20\%$. This difference can be explained by several factors. First of all, the analytical estimate, as mentioned, is an upper limit; a more detailed integration of the equation of motion gives a value several per cent lower. Secondly, the simulation does not treat the AGN wind hydrodynamically, so the outflow cavity is filled by a backflow from the shocked gas, further reducing the energy available to push it outwards. The total thermal energy of the gas contained in this hot bubble is $\sim 50\%$ of the total injected energy; reducing the energy injection by a factor of 2 in the analytical calculation brings the velocity estimate into agreement with the simulated one. Interestingly, the mass-weighted velocity, given in Table \ref{tab:sims}, is even lower, $\overline{v}_{\rm out} = 582$~km~s$^{-1}$. This happens because the velocity of the density peak is the highest that the gas reaches, while the velocity of the outer edge is the shockwave velocity, which does not correspond to real particle motion. Instead, there are particles getting accelerated towards $v_{\rm c.d.}$ but still have lower velocities, and they bring the average velocity down.

Figure \ref{fig:smooth_profiles} shows the radial profiles of the mass outflow rate (top) and the momentum and energy loading factors (middle and bottom, respectively) in the SmthAdiaL1.0 simulation at three times: $t = 0.2$~Myr (red), $t = 0.5$~Myr (green), and $t = 0.8$~Myr (blue). Each radial profile shows a single strong peak with a width increasing with time, from $\sim 50$~pc at $t=0.2$~Myr to almost $200$~pc at $t = 0.8$~Myr; again, this is expected and can be intuited from the density map. At all three times, the peak values are significantly higher than the analytical estimates; however, we must keep in mind that they represent values estimated in very narrow radial shells. The globally averaged values, shown by individual circles (for estimates using $\tau_{\rm AGN}$) and squares (for estimates using $\tau_{r/v}$), essentially agree with the analytical estimates, when accounting for the lower outflow velocity. The trend of values estimated using $\tau_{r/v}$ being lower than those estimated using the real outflow age is present throughout all simulations, as shown below (Sect. \ref{sec:time_evol}); we discuss the implications of this in Sect. \ref{sec:discuss_rv}. 

Outflow expansion in the SmthAdiaL0.3 simulation is qualitatively identical to that of SmthAdiaL1.0 (Fig. \ref{fig:smooth_density_L0.3}). Again, the velocity is lower than the analytical estimate ($\overline{v}_{\rm out} = 350$~km~s$^{-1}$, $v_{\rm c.d.} \simeq 400$~km~s$^{-1}$), for the same reasons as presented above.

\subsubsection{Smooth cooling simulations}


\begin{figure*}
\includegraphics[width=\textwidth, trim={0 10 0 10}, clip]{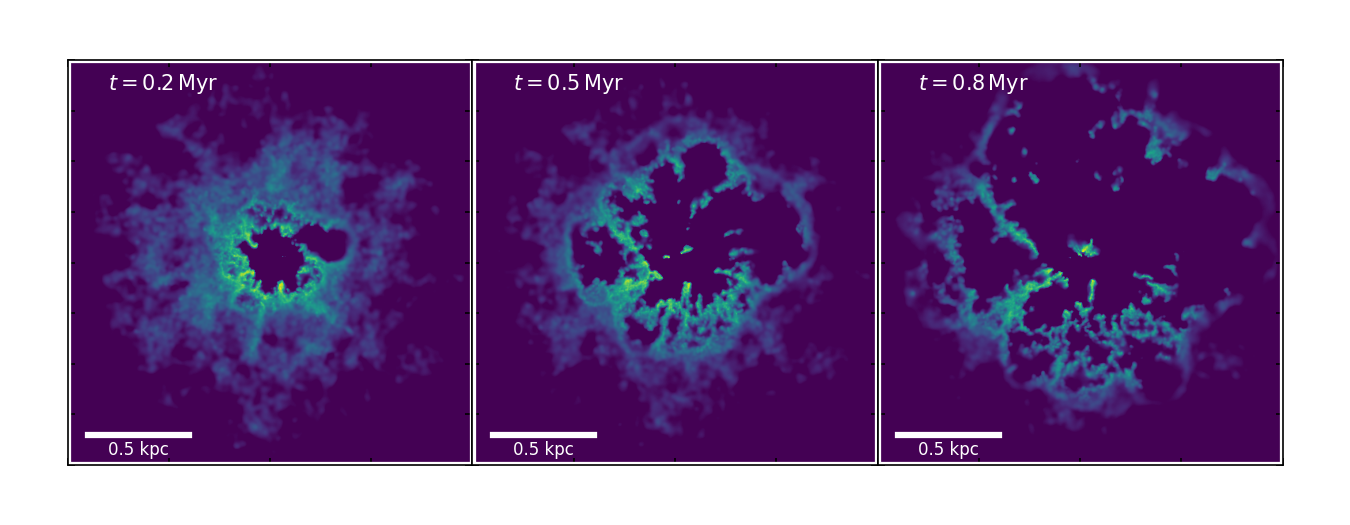}
\caption{Same as Fig. \ref{fig:smooth_density} but for the TurbCoolL1.0 simulation.}
\label{fig:turbulent_cooling_density}
\end{figure*}

\begin{figure}
\includegraphics[width=0.48\textwidth, trim={0 33 0 15}, clip]{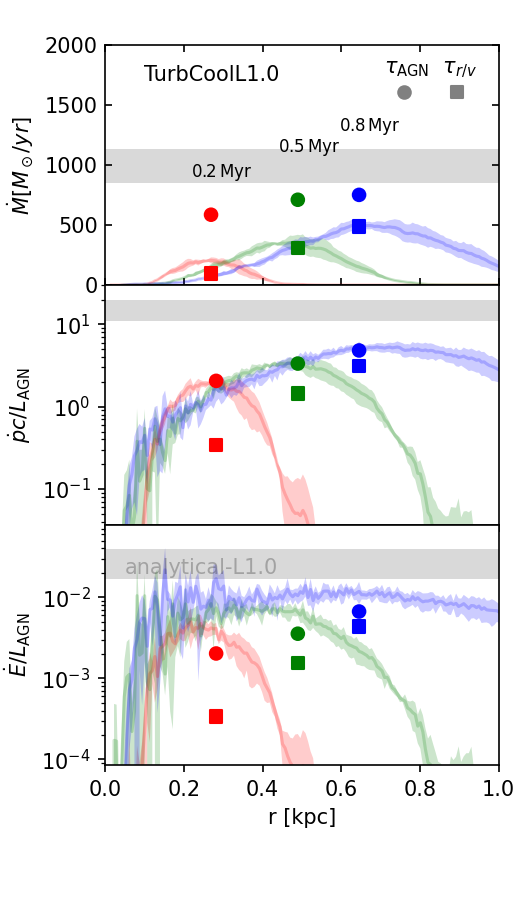} 
\caption{Same as Fig. \ref{fig:turbulent_profiles} but for the TurbCoolL1.0 simulation.}
\label{fig:turbulent_cooling_profiles}
\end{figure}

In Fig. \ref{fig:smooth_cooling_density} we show the density distribution of the outflow in simulation SmthCoolL1.0 as it evolves. Comparing these plots with Fig. \ref{fig:smooth_density}, we see that the outflow expands more slowly, with an average velocity $\sim 300$~km~s$^{-1}$, and the expanding region is confined to a narrower, $< 50$~pc thick, shell. The first difference occurs because the gas in the diffuse outflow cavity cools down and provides less push to the outflowing gas. The second difference occurs because the outflowing gas itself cools down and is compressed into a thinner shell by its own motion. The compressed shell also starts to fragment; this can be seen more clearly when we integrate the density throughout the simulation volume, as shown in Fig. \ref{fig:smooth_cooling_density_integrated}. Unlike in similar simulations performed in \citet{Nayakshin2014MNRAS}, the density variations across the shell are not very large: the ratio between top $5\%$ and bottom $5\%$ of density values does not exceed 100. This difference is due to the lack of self-gravity in our setup.

The same behaviour is evident when looking at the radial profiles (Fig. \ref{fig:smooth_cooling_profiles}). All three integrated parameters have lower values than in the adiabatic simulation. Although the peak values of $\dot{M}$ are only slightly lower, the globally integrated values are about twice as small as those in the adiabatic simulation. This happens because, as mentioned above, the outflow shell is much thinner than in the adiabatic simulation. As a result, even though the instantaneous mass flow rate through a given surface is large, the value integrated over the thickness of the shell is lower. The difference is starker for the other parameters and at later times: momentum loading is $\sim 3.5$ times lower at $t = 0.5$~Myr, while energy loading is $\sim 8$ times lower.  

Outflow expansion in simulation SmthCoolL0.3 proceeds along the same lines as in SmthCoolL1.0. The velocity is $\sim 3$ times lower than in SmthCoolL1.0, in contrast to the $\sim 1.5$ ratio between velocities in the two adiabatic simulations. This happens because, at a lower initial velocity, gas in the outflow is heated to lower temperatures where cooling is more efficient. So the fractional loss of energy in this simulation is greater than in the SmthCoolL1.0. The momentum loading factor is actually below unity, so such an outflow, if observed, would be seen as being driven by only the AGN wind momentum.

\subsubsection{Turbulent adiabatic simulations}

Turning now to the turbulent adiabatic simulations, we show the evolution of the density distribution in Fig. \ref{fig:turbulent_density}. While we can clearly see an expanding outflow, its shape is much more irregular, with different densities along the edge anti-correlated with the outflow distance from the AGN. By $t=0.5$~Myr, the outflow reaches the edge of the initial gas distribution at one point; later on, this becomes the site of a champagne outflow extending to very large distances. Radial profiles (Fig. \ref{fig:turbulent_profiles}) echo the same result: we no longer see a clear sharp peak, but a very widely distributed outflow region. In these plots, the shaded area around each line represents the standard deviation from the mean of the results in four simulations, identical except for a different random seed used to generate the turbulent velocity field (see Sect. \ref{sec:ic_gen}). The position of the maximum mass outflow rate (and, similarly, the maximum momentum and energy loading factors) moves with higher velocity than in the smooth simulations, extending out to almost $\sim 0.9$~kpc by $\sim 0.8$~Myr, giving a radial velocity of $\sim 1000$~km~s$^{-1}$. However, the large amounts of dense gas behind the peak reduce the average velocity to essentially the same as in the smooth simulation. Qualitatively, the same results are visible at other times and in both L1 and L0.3 simulations. Interestingly, the averaged values\footnote{These values are the same in simulations with different turbulence seeds, despite the considerable range covered by the shaded areas.} of all parameters are in remarkable agreement between the smooth and turbulent adiabatic simulations. At late times, when parts of the outflow expand beyond $R_{\rm out}$, the average velocity begins increasing, while mass and momentum rates start to decrease and the energy rate stays flat (see also Sect. \ref{sec:time_evol}). This, however, is an artefact of the initial conditions.

Simulation TurbAdiaL0.3 evolves essentially in the same way as TurbAdiaL1.0, except the blowout phase starts around $t = 0.8$~Myr. The radial profiles are also very similar, and the integrated quantities show basically the same evolution as the higher-luminosity run, with values of the loading factors being very similar to those in the corresponding smooth simulation.

\subsubsection{Turbulent cooling simulations}


\begin{figure}
\includegraphics[width=0.47\textwidth]{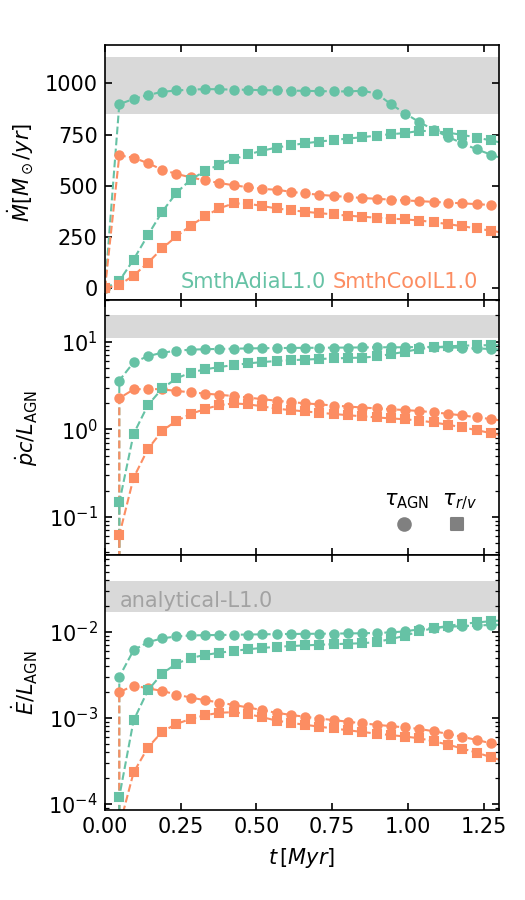} 
\caption{Evolution of the averaged mass outflow rate (top), momentum loading (middle), and energy loading factors (bottom) in the smooth L1.0 simulations with time. Teal and orange lines show estimates of the adiabatic and cooling simulations, respectively. Circles show estimates using $\tau_{\rm AGN}$, squares using $\tau_{r/v}$.}
\label{fig:smooth_evolution}
\end{figure}

\begin{figure}
\includegraphics[width=0.47\textwidth]{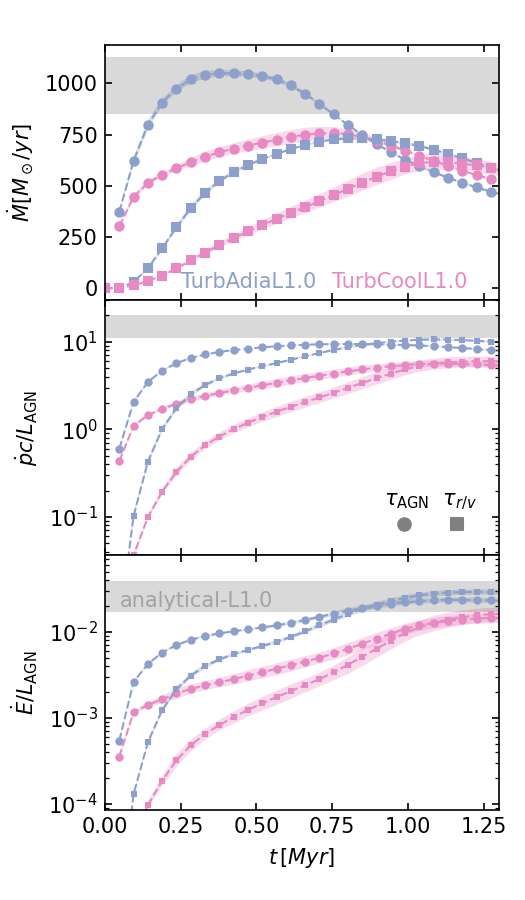} 
\caption{Same as Fig. \ref{fig:smooth_evolution} but for the turbulent simulations. Violet now reflects the adiabatic simulation, while magenta shows the cooling one.}
\label{fig:turbulent_evolution}
\end{figure}

Finally, we arrive at the TurbCoolL1.0 simulation, which has both a turbulent density distribution and gas cooling. Globally, the shape of the outflow (Fig. \ref{fig:turbulent_cooling_density}) is similar to that of the TurbAdiaL1.0 simulation, but the cooling outflow has much finer features and higher density contrasts, exceeding four orders of magnitude at distances $0.2-0.4$~kpc, compared to a ratio of $\sim 50$ outside the outflow. Multiple lobes are visible, separated by dense gas filaments. The average velocity is approximately twice lower than in the TurbAdiaL1.0 simulation but slightly higher than in the SmthCoolL1.0. This happens because the diffuse gas can be pushed away very efficiently, creating a tail of particles with very high radial velocities $\left(v > 1000\,{\rm km \,s}^{-1} \right)$ while a significant fraction of cold gas is still kept from collapsing to the centre very rapidly. The radial distributions of $\dot{M}_{\rm out}$ and loading factors (Fig. \ref{fig:turbulent_cooling_profiles}) emphasise this: outflowing material is distributed in a very broad shell, ranging from $\sim 0.25$~kpc to $>1$~kpc at $t = 0.8$~Myr, for example. The values of mass flow rate and momentum loading have visible peaks, while the energy loading factor is distributed almost uniformly with radius. The averaged values are a factor of $\sim 1.5-3$ lower than analytical estimates.

In simulation TurbCoolL0.3, the features of the TurbCoolL1.0 are even more pronounced. Density ratios are higher, velocity lower, and the radial distribution of the three parameters is almost uniform between the AGN and the outer edge of the outflow. Curiously, this means that the mass flow rate and momentum loading factor in the TurbCoolL0.3 simulation are lower than in the SmthCoolL0.3, reversing the trend seen when comparing other smooth and turbulent analogues.

\subsubsection{Time evolution of average outflow properties} \label{sec:time_evol}

In all simulations, outflow properties evolve with time. In the S-ad simulations, the $\tau_{\rm AGN}$-evaluations tend to settle to approximately constant values by $t = 0.2$~Myr (Fig. \ref{fig:smooth_evolution}, teal lines); evaluations using $\tau_{r/v}$ follow the same trend just take longer ($0.3-0.4$~Myr) to reach the plateau. Once the outflow clears the initial gas shell, the mass outflow rate begins to decrease roughly as $t^{-1}$, because the total outflowing mass remains the same, while $\tau_{\rm AGN}$ keeps increasing. Due to a corresponding increase in outflow velocity, the momentum loading factor hardly changes and energy loading increases somewhat. This happens because there is less $p{\rm d}V$ work needed to inflate the gas, so a larger fraction of the input energy is retained as kinetic energy of the outflow. In the cooling simulations (orange lines), all parameters continuously decrease as ever more energy is radiated away.  

In the T-ad simulations (Fig. \ref{fig:turbulent_evolution}, violet lines), the $\tau_{\rm AGN}$-evaluated mass outflow rate reaches a peak at $t \sim 0.3-0.5$~Myr and begins to decrease as soon as the outflow reaches the edge of the initial shell, eventually settling on the same $t^{-1}$ trend. The momentum loading factor stays essentially constant, while the energy loading increases as the outflow clears the shell, as in the smooth simulations. The $\tau_{r/v}$-evaluated values evolve with a delay, much like their smooth counterparts, and are still growing by $t \sim 0.7$~Myr. At this late stage, the velocity increases significantly as ever more of the outflow begins expanding into a vacuum. This leads to the $\tau_{r/v}$-estimated values overtaking the $\tau_{\rm AGN}$-estimated ones at $t \simeq 0.8$~Myr and to the energy loading factor increasing significantly. 

The T-c simulation evolution (Fig. \ref{fig:turbulent_evolution}, magenta lines) is remarkably different from the corresponding smooth simulation. Despite energy being radiated away, all parameters keep increasing with time up to $\sim 0.75$ ($1$)~Myr for the $\tau_{\rm AGN}$ ($\tau_{r/v}$) evaluation. At later times, as the outflow escapes the initial shell, mass outflow rates begin to decrease, but momentum loading stays approximately constant, while energy loading keeps increasing. This happens because while outflowing gas has a broad temperature distribution, meaning that cold gas contributes to a significant mass outflow rate, the energy rate (and energy loading factor) is completely dominated by the hot gas, which cools inefficiently. We present this in more detail in Sect. \ref{sec:gas_phase}.

\begin{figure}
\includegraphics[width=0.47\textwidth]{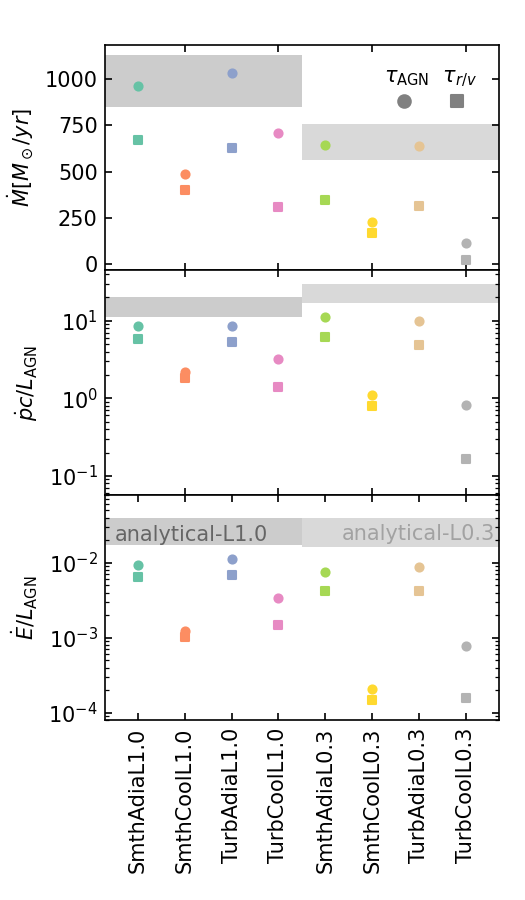} 
\caption{Values of the mass outflow rate (top), momentum loading (middle), and energy loading factors (bottom) in all four L1 simulations at $t = 0.7$~Myr. The grey area is the analytical prediction.}
\label{fig:summary_plot}
\end{figure}

\subsubsection{Global effect of turbulence and cooling} \label{sec:profiles_summary}

For ease of comparison, at a glance, of the effects of cooling, turbulence and both combined on the outflow properties, we plot the average values of the relevant parameters at $t = 0.5$~Myr in Fig. \ref{fig:summary_plot}. As in earlier plots, circles show values obtained using $\tau_{\rm AGN}$, while squares show those obtained using $\tau_{r/v}$. As noted above, there is virtually no difference between the SmthAdiaL1.0 and TurbAdiaL1.0 simulations or between the SmthAdiaL0.3 and TurbAdiaL0.3 ones; that is to say, the addition of turbulence has no detrimental effect on the coupling efficiency of AGN wind momentum and energy to the surrounding ISM. Even though the energy is distributed in a much wider shell in the turbulent simulations, the total energy injection remains the same. In fact, turbulence increases the kinetic energy slightly, because there is more diffuse high-velocity gas in those simulations.

When we compare the results of simulations with cooling, a curious pattern emerges: the TurbCoolL1.0 simulation has significantly higher $\tau_{\rm AGN}$-evaluated values of all parameters than SmthCoolL1.0. Cooling in the smooth simulation is particularly effective because all of the outflowing material is compressed into a thin and dense shell, while in the turbulent simulations, a lot of the outflowing material remains diffuse and so cools inefficiently. However, in the L0.3 simulations, the trend is reversed for mass and momentum outflow rates: the addition of turbulence leads to a decrease because lower AGN luminosity leads to less hot gas being produced and cooling is overall more efficient. 

Overall, the combined effect of turbulence and cooling, when compared with the smooth adiabatic setup, results in a reduction of the mass outflow rate by $\sim25\%$ ($\sim80\%$), the momentum loading factor by $\sim60\%$ ($\sim93\%$) and the energy loading factor by $\sim63\%$ ($\sim90\%$) in the L1 (L0.3) simulations at $t = 0.5$~Myr. The difference in temporal evolution means that the reduction is greater at earlier times but becomes less significant as the outflow evolves.

\subsection{Gas phase distribution} \label{sec:gas_phase}

\begin{figure}
\includegraphics[width=0.47\textwidth, trim={0 6 0 15}, clip]{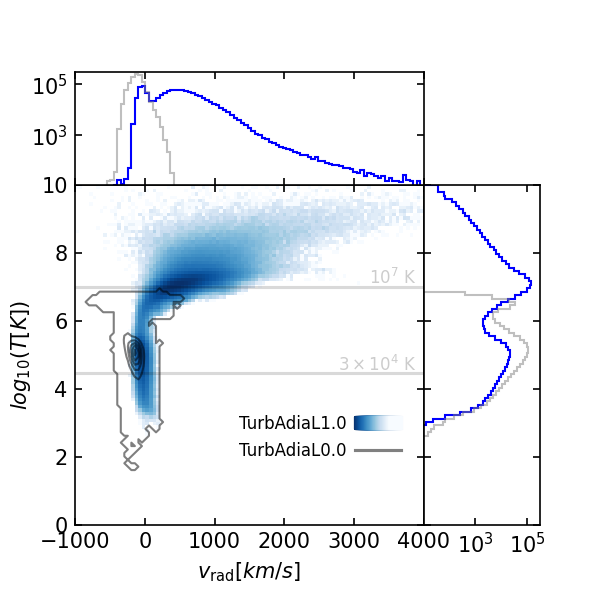} 
\includegraphics[width=0.47\textwidth, trim={0 6 0 15}, clip]{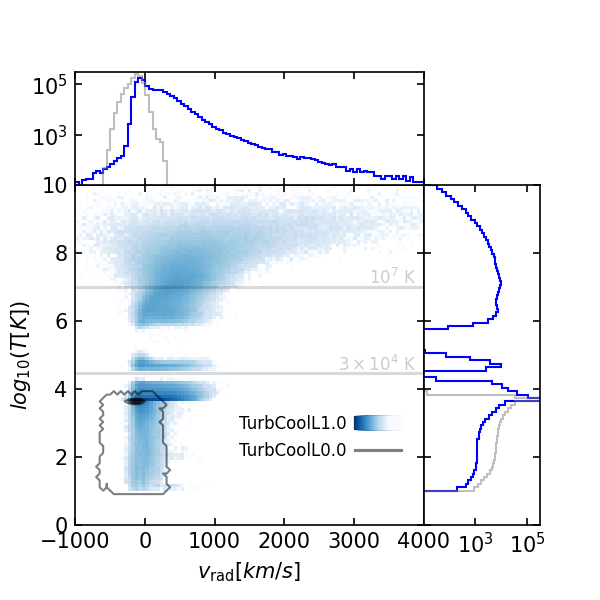} 
\caption{Two-dimensional histograms of gas temperature (vertical axis) against radial velocity (horizontal axis) in the turbulent simulations. Grey contours show corresponding control simulation data. Grey horizontal lines show the criteria for classifying gas as cold, intermediate, or hot.}
\label{fig:temp_histograms}
\end{figure}

\begin{table*}
  \centering
  \def\arraystretch{1.2}
  \caption{Properties of multi-phase gas in the turbulent simulations with cooling.}
  \begin{tabular}{@{}lccccccc@{}}    
    \toprule
    Run & phase & temperature range &  $v_{\rm out} $ & $\dot{M}_{\rm out}$ & $\dot{p}_{\rm out} c/L_{\rm AGN}$ & $\dot{E}_{\rm k, out}/L_{\rm AGN}$ & $f_{\rm esc}$\\  
    & & & $\left(\text{km s}^{-1}\right)$ & $\left(\msun \text{ yr}^{-1}\right)$ & & $\left(\times 10^{-3}\right)$ & $\%$ \\
    \midrule
    TurbCoolL0.3 & cold & $< 3\times10^4$~K & 91.4 & 92 (11.1) & 0.42 (0.05) & 0.13 (0.02) & 0.0012 \\
    TurbCoolL0.3 & intermediate & $3\times10^4$~K - $10^7$~K & 221 & 9.2 (4.9) & 0.1 (0.05) & 0.06 (0.03) & 0.54 \\
    TurbCoolL0.3 & hot & $> 10^7$~K &  603 & 15 (33) & 0.45 (0.99) & 0.95 (2.1) & 21 \\
    \hline
    TurbCoolL1.0 & cold & $< 3\times10^4$~K & 209 & 495 (160) & 1.55 (0.50) & 0.83 (0.27) & 0.19 \\
    TurbCoolL1.0 & intermediate& $3\times10^4$~K - $10^7$~K & 316 & 93 (35) & 0.44 (0.16) & 0.33 (0.12) & 3.4 \\
    TurbCoolL1.0 & hot& $> 10^7$~K & 691 & 122 (114) & 1.27 (1.19) & 2.42 (2.26) & 34 \\
    \hline
    \vspace{0.1cm}
  \end{tabular}
  \vspace{0.3cm}
  \begin{minipage}{\textwidth}
    \small
    \textbf{Notes:} The first column shows the simulation name, the second and third give the name and temperature range of the gas phase. The subsequent columns show the results: mass-weighted outflow velocity, time-averaged mass outflow rate, momentum and energy loading factors, and the fraction of gas reaching escape velocity (eq. \ref{eq:vesc}). All numbers are estimated as in Table \ref{tab:sims}.
  \end{minipage}
  \label{tab:phases}
\end{table*}

To better understand how the AGN feedback energy couples to the gas and to connect our results to observations of multi-phase outflows, we considered the major outflow properties of cold, warm, and hot gas in the simulations. We mainly present the results for the turbulent simulations with cooling, with a few comments regarding the others when necessary.

In Fig. \ref{fig:temp_histograms} we show two-dimensional histograms of gas temperature and radial velocity in four turbulent simulations: TurbAdiaL1.0 (top, blue points and lines), TurbCoolL1.0 (bottom) and their corresponding control simulations (grey contours and lines). Gas with temperatures below $10^4$~K exists essentially only in the cooling simulations; we chose a significant drop in particle numbers at $T = 3\times10^4$~K in the TurbCoolL1.0 simulation to delineate cold gas. Gas with temperatures above $10^7$~K exists only in simulations with an AGN; it represents significantly shock-heated gas and we used this temperature to distinguish hot gas. We show these limits with grey horizontal lines in the histograms.

\begin{figure}
\includegraphics[width=0.47\textwidth]{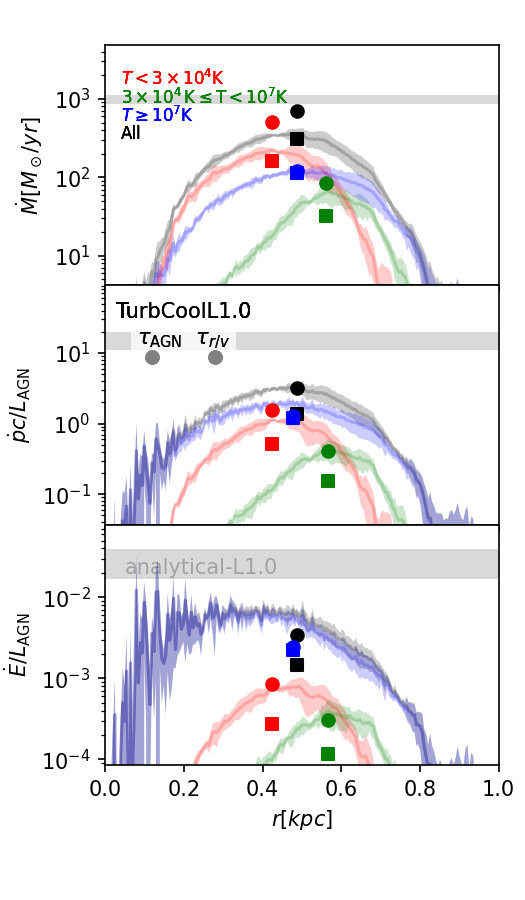} 
\caption{Radial profiles of the mass outflow rate (top), momentum loading (middle), and energy loading factors (bottom) of cold (red lines and points), warm (green), and hot (blue) gas in simulation TurbCoolL1.0 at $t= 0.5$~Myr. The total values are shown in black.}
\label{fig:tc_profiles_by_phase}
\end{figure}

We plot the mass outflow rates and the momentum and energy loading factors of gas of the three phases in the TurbCoolL1.0 simulation at $t=0.5$~Myr in Fig. \ref{fig:tc_profiles_by_phase} and give the averaged values in Table \ref{tab:phases}. The cold gas dominates the mass outflow rate at distances $0.2-0.6$~kpc from the nucleus, while hot gas becomes dominant further out. The mass budget is also dominated by cold gas, which comprises $\sim 70\%$ of the total, followed by hot gas at $\sim 20\%$. As expected, hot gas becomes more important when one considers momentum and especially energy loading: the momentum loading factors of hot and cold gas are comparable, while the energy loading factor of the hot gas is several times higher than that of the cold.

The radial velocities of gas of different phases show remarkable differences. In the adiabatic simulation, cold gas has a narrow velocity distribution peaking around zero, with maximum velocities $v_{\rm c, max} \simeq \pm 2 \sigma_{\rm b}$. This represents gas that the outflow has not yet reached. Gas above $10^6$~K shows increased radial velocities, but warm gas has maximum velocities $\sim 1500$~km~s$^{-1}$. Hot gas contains material that has escaped from the initial gas shell and so moves with the highest velocities exceeding even $4000$~km~s$^{-1}$. In the TurbCoolL1.0 simulation, there is cold outflowing gas as well, but its velocity does not exceed  $1000$~km~s$^{-1}$; the same is true for warm gas. Conversely, the hot gas still has very large radial velocities. It is interesting to compare gas velocities with the escape velocity from the simulated galaxy; for a singular isothermal sphere, the escape velocity to infinity is not well defined; instead, we chose $r_{\rm esc} = 100$~kpc as a proxy. Then the escape velocity becomes
\begin{equation}\label{eq:vesc}
    v_{\rm esc} = 2 \sigma_{\rm b} \left(1 + {\rm ln} \left(r_{\rm esc} / r\right)\right)^{1/2},
\end{equation}
where $r$ is the current radial coordinate of the particle. For initial distances $0.1 < r <1$~kpc from the nucleus, the escape velocity range is $4.7 \sigma_{\rm b} < v_{\rm esc} < 5.6 \sigma_{\rm b}$ or  $667$~km~s$^{-1} < v_{\rm esc} < 795$~km~s$^{-1}$. The fraction of all gas that exceeds this velocity at $0.5$~Myr is $3.6\%$; this number increases to $12\%$ by $0.8$~Myr. However, the fraction of escaping cold gas is only $0.19\%$, increasing to $0.77\%$. This is consistent with observational results showing that only a small fraction of outflowing material has enough kinetic energy to escape the host galaxy \citep{Fluetsch2019MNRAS}. On the other hand, more than one-third of hot gas is escaping at $0.5$~Myr, increasing to half by $0.8$~Myr, reinforcing the point that most of the injected AGN energy is carried away by the hot phase.

\begin{figure}
\includegraphics[width=0.47\textwidth]{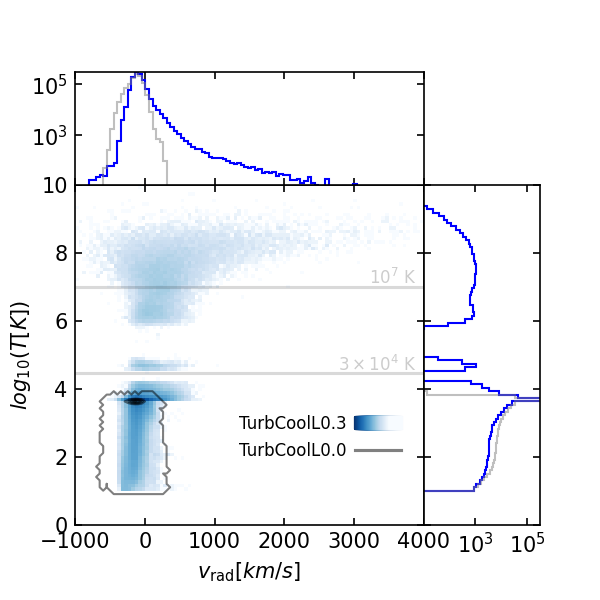} 
\caption{Same as Fig. \ref{fig:temp_histograms} but for the TurbCoolL0.3 simulation.}
\label{fig:temp_histograms_l03}
\end{figure}

The distribution of gas temperatures and velocities in the TurbCoolL0.3 simulation is qualitatively very similar to that in TurbCoolL1.0 (Fig. \ref{fig:temp_histograms_l03}). There is less hot gas, but it still dominates the energy budget. The mean velocities, especially of the cold component, are lower. However, when we look at the velocities required to escape the galaxy, the picture is very different. Only $0.28\%$ of gas is escaping at $t = 0.5$~Myr, 10 times less than in the higher-luminosity simulation; this number stays essentially the same at $0.8$~Myr, suggesting that further energy input is not enough to unbind any more of the gas. When we consider only the cold gas, the escaping fraction is $1.2\times 10^{-5}$, two orders of magnitude lower than in TurbCoolL1.0; by $0.8$~Myr, it has increased to $1.1 \times 10^{-4}$, still a factor of $\sim 70$ lower than in the higher luminosity run. The fraction of escaping warm gas also increases by almost an order of magnitude, from $0.54\%$ to $4.7\%$, while the fraction of hot gas increases only a little, from $21\%$ to $27\%$. This trend of escaping gas fraction with luminosity suggests that there is a threshold, probably around $L = L_{\rm Edd}$, where even cold gas can be efficiently pushed away by an outflow. Such behaviour agrees with previous simulations \citep{Zubovas2014MNRASb} and represents a way to establish the $M-\sigma$ relation in the absence of efficient cooling of the shocked AGN wind \citep{Faucher2012MNRASb}. We intend to explore the dependence of outflow phases on AGN luminosity and the details of the cooling prescription in a future publication.

\section{Discussion} \label{sec:discuss}

\subsection{Comparison with real outflows} \label{sec:discuss_multiphase}

In our turbulent simulations with cooling, the cold gas phase dominates the mass budget, comprising $80\%$ ($70\%$) of the mass outflow rate in the L0.3 (L1) simulation. The momentum loading factors of cold gas are comparable to those of the hot, while the energy loading is dominated by the hot component (see Table \ref{tab:phases}). Furthermore, the hot phase moves with 3-6 times higher velocity and so extends further out - the average radius is $\sim 15\%$ higher at $t = 0.5$~Myr (Fig. \ref{fig:tc_profiles_by_phase}). Looking at the compilations of outflow data from \citet{Fiore2017AA}, \citet{Fluetsch2019MNRAS}, and \citet{Lutz2020AA}, we see that our results qualitatively agree with real outflow parameters. Quantitatively, some discrepancies are present. Real ionised outflows typically have higher velocities, $\overline{v}_{\rm ion} \sim 1100$~km~s$^{-1}$, and so do molecular ones, $\overline{v}_{\rm mol} \sim 450$~km~s$^{-1}$. Both velocities correlate with AGN luminosity; at $L_{\rm AGN} \sim \left(0.3-1\right)\times 10^{46}$~erg~s$^{-1}$, the mean ionised outflow velocity is the same, while that of molecular outflows increases to $\sim 650$~km~s$^{-1}$. Molecular outflows are observed to have on average higher mass outflow rates at $L_{\rm AGN} \sim 10^{46}$~erg~s$^{-1}$ and momentum loading factors, while the energy loading factors are comparable between the two phases. On the other hand, in the few objects where several outflow phases have been detected simultaneously, the molecular phase is dominant both in terms of mass and energy \citep{Fluetsch2021MNRAS}. Finally, ionised outflows tend to be observed at higher distances of several kpc, while molecular gas is more commonly found in the central kiloparsec, even at $L_{\rm AGN} \sim 10^{46}$~erg~s$^{-1}$. So our simulations are underestimating all outflow velocities, but the cold gas velocity is underestimated more; simultaneously, the radial extent of the hot outflow component is underestimated while its mass content is (slightly) overestimated. The reason for the discrepancy in the hot component properties may be the lack of hydrodynamic AGN wind. Because of this, the hot gas phase in our simulations extends inwards towards the AGN, filling the region that, in reality, is taken up by the shocked wind. Furthermore, we did not model the halo gas, which is comparatively very diffuse and would contribute to the hot phase of the outflow. The lack of rapidly moving cold gas may be a consequence of our assumption that the gas is optically thin and the neglect of gas self-gravity. This makes cooling less efficient and precludes the precipitation of cold dense clumps from outflowing hot gas \citep{Nayakshin2012MNRASb, Zubovas2013MNRAS}.

Our simulated outflow properties generally have a rather clear dependence on AGN luminosity. In particular, mean outflow velocities in L1.0 simulations are 1.6-2 times higher than in L0.3, except for the SmthCool simulations, where the ratio is 3.16. The mass outflow rates also differ by a factor of 1.5-2.1, except for TurbCool, where the ratio is an exceptional 6.1. This can be compared with the analytical expectation $v_{\rm out} \propto L_{\rm AGN}^{1/3}$, which gives $v_{\rm L1.0}/v_{\rm L0.3} \simeq 1.5$. The mass outflow rate ratio in the adiabatic simulations agrees with the expectation, while the velocity ratio is only slightly higher, most likely due to the presence of hot gas filling the outflow cavity, which is fractionally more important in the L0.3 runs. When cooling is introduced, the lower-luminosity AGN heats the gas to lower temperatures, leading to more efficient cooling and a greater difference between the analytical estimate and simulated properties. 

Observationally, outflow velocity scales roughly as $L^{0.16-0.29}$ \citep{Fiore2017AA}; this relationship is flatter than the analytical estimate and much more discrepant with our results. The mass outflow rate, on the other hand, has a steeper dependence on luminosity: $\dot{M}_{\rm out} \propto L^{0.76-1.29}$ \citep{Fiore2017AA}; this is also discrepant with our results, as it predicts a difference by a factor of $2.5-4.7$. Understanding why the two relationships differ so much helps explain their difference from the simulated results too. In our analytical estimate,
\begin{equation}
    v_{\rm out} \propto f_{\rm g}^{-1/3} \sigma_{\rm b}^{-2/3} L_{\rm AGN}^{1/3},
\end{equation}
while the mass outflow rate
\begin{equation}
    \dot{M}_{\rm out} \propto f_{\rm g} \sigma_{\rm b}^2 v_{\rm out} \propto f_{\rm g}^{2/3} \sigma_{\rm b}^{4/3} L_{\rm AGN}^{1/3}.
\end{equation}
In our earlier calculations, we assumed that the only difference between the high- and low-$L_{\rm AGN}$ systems is the AGN luminosity. In reality, the gas velocity dispersion, which is closely related to the galaxy and SMBH masses, has a systematic dependence on $L_{\rm AGN}$, and the gas fraction may also have such a dependence. Assuming that $\sigma_{\rm b} \propto M_{\rm BH}^{1/4} \propto \left(L_{\rm AGN}/l\right)^{1/4}$, where $l \equiv L_{\rm AGN} / L_{\rm Edd}$ is the Eddington ratio, we find
\begin{equation}
    v_{\rm out} \propto f_{\rm g}^{-1/3} L_{\rm AGN}^{1/6},
\end{equation}
while the mass outflow rate
\begin{equation}
    \dot{M}_{\rm out} \propto f_{\rm g}^{2/3} L_{\rm AGN}^{2/3}.
\end{equation}
We see that even without any correlation between $f_{\rm g}$ and $L_{\rm AGN}$, we recover almost exactly the observed correlations $v_{\rm out} \propto L_{\rm AGN}^{0.16}$ and $\dot{M}_{\rm out} \propto L_{\rm AGN}^{0.66}$. So the observed correlations can be explained if the different AGN have the same distribution of Eddington ratios, and the luminosity difference arises from the difference in SMBH (and host galaxy) masses, an aspect we did not cover in our simulations. As the number of outflow observations with known SMBH masses increases, it will be interesting to check whether the relationship between outflows in galaxies differing only in AGN Eddington ratio follows our simulated results (and analytical predictions) more closely than the whole population.




\subsection{Determining outflow properties from observations} \label{sec:discuss_rv}

\begin{figure}
\includegraphics[width=0.47\textwidth]{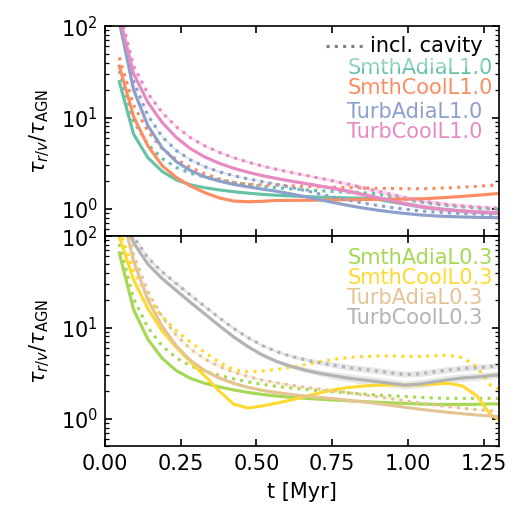} 
\caption{Ratio of the two outflow age estimates, $\tau_{r/v}/\tau_{\rm AGN}$, as a function of time in all simulations. Dotted lines are calculated without accounting for the presence of the initial cavity in the gas distribution.}
\label{fig:timescale_estimates}
\end{figure}

Throughout our simulations, the average outflow properties estimated using $\tau_{r/v}$ are almost always higher than those estimated using $\tau_{\rm AGN}$. In Fig. \ref{fig:timescale_estimates} we plot the ratio of these two timescale estimates as a function of time for all eight non-control simulations. The solid lines correspond to the actual $\tau_{r/v}$ used when estimating outflow parameters, while dotted lines correspond to the ratio $\overline{R}_{\rm out} / \overline{v}_{\rm out}$ (i.e. ignoring the presence of the central cavity at the start of the simulation). We see that initially, $\tau_{r/v}$ can be more than an order of magnitude higher than $\tau_{\rm AGN}$ (i.e. the real age of the outflow), and generally remains higher throughout the simulations. Only when the outflow breaks out of the initial gas shell, at $t > 0.8$~Myr in simulations SmthAdiaL1.0, TurbAdiaL1.0 and TurbCoolL1.0, does the ratio drop below unity. This means that using the commonly adopted equation $\dot{M}_{\rm out} = M_{\rm out} v_{\rm out} / R_{\rm out}$ provides mass outflow estimates that are up to several times too small. The discrepancy is higher when the AGN luminosity is lower and when the outflow is younger (i.e. closer to the nucleus). This estimate is sometimes multiplied by a factor of a few, which would bring it into an agreement with the real value; however, this multiplication is usually done when trying to account for different outflow geometries \citep[e.g.][]{Cicone2014AA, Gonzalez2017ApJ}. We plan to explore the relation between these outflow estimates and geometry in a future paper.

\subsection{Establishing the $M-\sigma$ relation}

One of the reasons why cooling is much more efficient in the low-luminosity turbulent simulations is the presence of dense gas clumps. While the clumps get slightly heated by the outflow, they cool down quickly and remain resilient to the feedback. They can subsequently fall onto the SMBH and continue feeding and growing it. By $t = 0.5$~Myr, $\sim 85\%$ of the cold gas is falling inwards and this number is essentially unchanged by $t = 0.8$~Myr. In the high-luminosity simulation, the total mass of such clumps is much smaller - $\sim 45\%$ at $t = 0.5$~Myr and only $\sim 13\%$ at $t = 0.8$~Myr. While we do not have simulations with higher luminosity, we expect this trend to continue: cold gas clumps will be pushed away, heated and dispersed ever more efficiently as the AGN luminosity increases. This confirms a way to establish the $M-\sigma$ relation without having explicitly momentum-driven outflows \citep{King2010MNRASa}, as predicted by considering the two-temperature nature of the shocked wind plasma \citep{Faucher2012MNRASb}. At low luminosities, the cold dense clumps, resilient to heating and evaporation, are not pushed by the high-energy outflow and continue feeding the SMBH. As the luminosity increases, the wind momentum becomes high enough to push the dense clumps away, as is happening in the TurbCoolL1.0 simulation. Then the SMBH feeding stops and its mass is established as given by the momentum-driven wind formalism \citep{King2010MNRASa}. This has been shown before in idealised simulations with global density gradients \citep{Zubovas2014MNRASb}, but here we see the same process working on smaller density inhomogeneities.

\subsection{Implementation of AGN feedback in numerical simulations}

As noted in Sect. \ref{sec:results_S_ad}, the lack of hydrodynamic realisation of an AGN wind leads to a reduction in outflow energy by a factor of $\sim 2$. The precise reduction is lower in simulations with turbulence and in simulations with higher luminosity, but these dependences appear weak; there may also be a dependence on gas density. This issue is almost certainly important for cosmological simulations, which typically implement AGN feedback as a simple injection of kinetic and/or thermal energy into the surrounding gas \citep[e.g.][]{Booth2009MNRAS, Vogelsberger2014MNRAS, Tremmel2017MNRAS, Dave2019MNRAS, Nelson2019MNRAS}. As the outflow bubble expands in those simulations, the lack of pressure from the presumably extremely hot shocked wind or a thermalising jet leads to a backflow, leaving less energy to push the gas outwards, much like in our simulations.

We envision a few ways to mitigate this issue. The numerically most straightforward way is to double the formal AGN feedback efficiency. However, this would merely mask the problem and would not account for the possible influence of gas density, outflow size and shape. A more realistic solution would be to add a pressure term to the SMBH particle, with the value of this pressure directly proportional to the total injected AGN feedback energy and inversely proportional to the volume of the cavity formed by the shocked wind or jet. The injected energy should be modified by cooling, which becomes important once the AGN luminosity decreases and the wind energy is no longer maintained. The volume of the cavity can be approximated by using the distance to the gas particles neighbouring the SMBH in SPH simulations, and by considering a density threshold in grid-based ones. A drawback of this approach is that the pressure is necessarily isotropic and cannot account for such effects as outflow breakout through low-density channels. A more detailed and anisotropic solution would be to track wind pressure using the same grid that was used for feedback injection (see Sect. \ref{sec:injection}). That way, shocked wind properties can be tracked in each direction independently, providing a continuous push to the gas particles at the inner edge of the outflow cavity.

Of course, the best way of eliminating the problem is to actually treat the wind hydrodynamically. This has been done in AREPO moving-mesh simulations \citep{Costa2020MNRAS}, where the quasi-relativistic wind is injected into cells neighbouring the SMBH particle, as well as for jet feedback in idealised \citep{Bourne2017MNRAS, TalbotEtAl22, Talbot2024MNRAS, Ehlert2023MNRAS} and cosmological \citep{Bourne2019MNRAS, Bourne2021MNRAS} simulations, by using novel refinement techniques. In general, improvements of numerical resolution around SMBHs \citep{Curtis2015MNRAS, Hopkins2024OJAp} can pave the way to detailed treatments of extreme gas in their vicinity and lead to better prescriptions for making feedback and its effects more realistic \citep{CurtisSijacki16, Bourne2017MNRAS, Koudmani2019MNRAS, Talbot2021MNRAS}.

\begin{figure}
\includegraphics[width=0.47\textwidth]{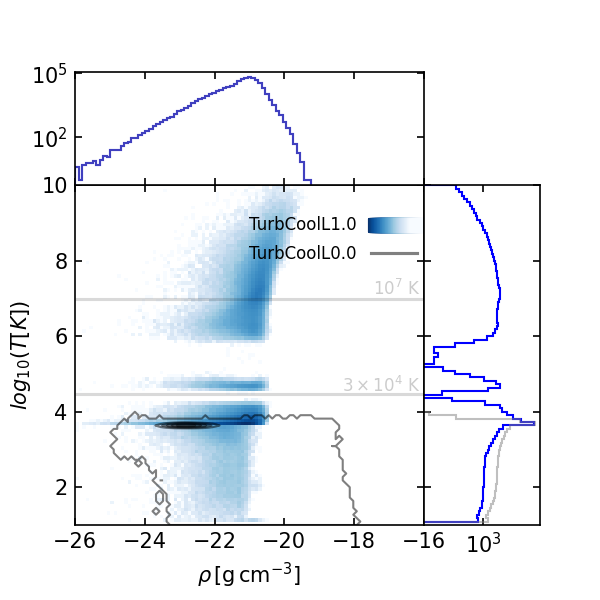} 
\caption{Same as Fig. \ref{fig:temp_histograms} but showing the relationship between gas temperature and density.}
\label{fig:phase_diagram_TurbCoolL1}
\end{figure}

Another aspect of our results relevant to larger-scale simulations is connected to the spatial resolution. We showed that the uneven density distribution arising due to turbulence interacts with cooling in a non-trivial way. Depending on the AGN luminosity, the clumpiness of gas can either enhance or suppress the mass and momentum rates of the outflow. The connection relies on the gas temperature distribution: diffuse hot gas cools inefficiently, while dense gas can radiate away injected energy very rapidly. In addition to AGN luminosity, the connection almost certainly depends on average gas density and level of turbulence, which determines the typical ratio of densities at a given distance from the nucleus. These results echo the conclusion of \citet{Bourne2015MNRAS} that low-resolution simulations are better at destroying galaxies via AGN feedback because a more even density distribution does not allow injected energy to escape. Cosmological simulations usually cannot resolve gas with density exceeding $n_{\rm H} \sim 100$~cm$^{-3}$ \citep[e.g.][]{Nelson2019MNRAS}, corresponding to $\rho \sim 10^{-22}$~g~cm$^{-3}$, which is lower than the average gas density on the inside of our simulated shell at the start of the simulations. As the simulation evolves, most of the cold gas, but also some warm and even hot gas attain higher densities (see Fig. \ref{fig:phase_diagram_TurbCoolL1}). As a result, cosmological simulations may overpredict the amount of hot gas in outflows and underpredict the cooling rate of outflowing gas. Even dedicated smaller-scale simulations may be under-predicting the mass of cold dense gas \citep[e.g.][]{Nuza2014MNRAS, Valentini2017MNRAS}. At the same time, cosmological simulations generally do not resolve gas of very low density either, over-predicting the cooling rate. There are two common approaches to mitigating this issue. The first, preventing gas particles that receive AGN feedback injection from cooling for a while \citep[e.g.][]{Tremmel2017MNRAS, Tremmel2019MNRAS}, does not capture the nuances of this process, leading to an unrealistic distribution of different gas phases. The second method, accumulating AGN feedback energy for a significant period of time \citep[e.g. 25 Myr; see][]{Henden2018MNRAS} before injecting it into the gas in one explosive event, misses the gradual development of outflows. Potentially the best way to reduce this problem is to use multi-phase particles \citep{Springel2003MNRAS, Murante2010MNRAS, Valentini2017MNRAS}, where each particle (or cell in a grid-based method) is assumed to contain gas of two or three phases, each with distinct temperature and density. Furthermore, improved numerical resolution around shocks and other density discontinuities \citep{Bennett2020MNRAS} can help us better track the evolution of multi-phase gas within an outflow.

\subsection{Model caveats} \label{sec:discuss_caveats}

In this study we are interested in capturing the effect and interplay of turbulence and cooling of AGN wind-driven outflows. For this reason, our simulations are heavily idealised. While their results can be used to interpret real outflow data and improve large-scale numerical simulations (see the rest of this section), there are many possible improvements.

First of all, in real galaxies, even the central spheroid usually has some angular momentum, which facilitates outflow escape via the polar directions \citep{Zubovas2012MNRASa, Zubovas2014MNRASb, CurtisSijacki16}. This is enhanced further by the presence of a disc. Another effect that may be collimating the large-scale feedback are small-scale anisotropies \citep[i.e. the conical geometry of the AGN wind;][]{Proga2004ApJ, Nardini2015Sci, Luminari2018A&A}. The wind geometry in any particular source is highly uncertain and so would be another free parameter in our simulations. In general, the polar direction of the accretion disc need not line up with the polar direction of the galaxy \citep[as evidenced by the directions of observed AGN jets;][]{Kinney2000ApJ} and the interplay between collimation on different spatial scales can lead to further complexities in outflow geometry.

We neglected gas self-gravity in these simulations, which precludes the formation of very dense clumps and stars. Star formation consumes some material and so reduces the mass outflow rate; simultaneously, stellar feedback may combine with AGN feedback to enhance the energy of outflows.

The cooling function we adopted is also simplified by assuming constant Solar metallicity of the gas. It is well known that more metal-rich gas cools more quickly \citep{Costa2015MNRAS} and so will preferentially precipitate out of the outflow. This should lead to the formation of high velocity, high density cold metal-rich gas \citep{Zubovas2014MNRASa, Richings2018MNRAS, Richings2018MNRASb}. The assumption that gas is optically thin also leads to unrealistically inefficient cooling of dense clumps. These clumps may efficiently transport metal-rich gas to the galactic outskirts, the circumgalactic or even intergalactic medium, affecting the chemical evolution of the galaxy and its environment.

Finally, our assumption of constant AGN luminosity is unrealistic. Both observations \citep{Schawinski2015MNRAS} and analytical arguments \citep{King2015MNRAS} suggest that individual AGN episodes should last only $\sim 0.1$ Myr, although they may be clustered into longer phases of enhanced activity \citep{Hopkins2007ApJ, Zubovas2022MNRAS}. This finding is corroborated by numerical simulations of realistic accretion of interstellar gas clouds \citep{Alig2011MNRAS, Tartenas2020MNRAS, Tartenas2022MNRAS}. A `flickering' AGN would interact with gas cooling in a non-linear way, because the cooling timescale of some gas is shorter than the expected downtime between successive AGN episodes, while the hot gas would stay hot throughout.

We plan to address most of these caveats in future works, gradually building up a realistic picture of SMBH accretion and feedback from kiloparsec down to sub-parsec scales. This will both enhance our understanding of real outflows and suggest ways to improve the sub-resolution prescriptions used in large-scale numerical simulations.

\section{Summary} \label{sec:sum}

In this paper we have presented the results of hydrodynamic simulations of AGN wind-driven outflows in idealised galaxy bulges, focusing on the effects of turbulence, cooling, and their influence on the major properties of the outflows: the mass outflow rate and the momentum and energy loading factors. Our main results are the following:
\begin{itemize}
    \item Simulations of smoothly distributed gas under the adiabatic equation of state produce spherically symmetric outflows with properties in general agreement with analytical expectations, except that a significant fraction of the injected energy remains in the outflow cavity where the shocked AGN wind would be in reality; this leads to outflows being slower and having lower momentum and energy than predicted.
    \item The addition of turbulence has almost no effect on the coupling between AGN wind and the gas; in fact, the outflows become slightly more energetic in the turbulent simulations.
    \item Cooling, on the other hand, has a significant effect, reducing the outflow energy by one to two orders of magnitude in the simulations with smoothly distributed gas and by up to one order of magnitude in the turbulent simulations.
    \item The interplay between cooling and turbulence is not straightforward and depends on AGN luminosity: in simulations with $L_{\rm AGN} = L_{\rm Edd}$, turbulence mitigates cooling by allowing for a large amount of gas to be heated to very high temperatures leading to inefficient cooling, while in simulations with $L_{\rm AGN} = 0.3 L_{\rm Edd}$, turbulence enhances the effect of cooling on the mass and momentum rates by creating dense gas clumps that are resilient to feedback and can maintain their density by cooling. The destruction of such clumps in the higher luminosity simulations can lead to the establishment of the $M-\sigma$ relation.
    \item In the most realistic simulation -- with both turbulence and cooling -- cold gas dominates the mass outflow rate, cold and hot gas have similar momentum rates, and the energy rate is dominated by the hot gas. The hot gas has a significantly higher velocity and is distributed farther out than the cold. This agrees qualitatively with the properties of observed outflows.
    \item As outflows evolve and break out from the initial gas shell, their velocity increases and the mass outflow rate decreases; the combined effect leads to an increase in the kinetic energy rate.
    \item Estimates of average mass outflow rates obtained using the common observational prescription $\dot{M}_{\rm out} = M_{\rm out} v_{\rm out} R_{\rm out}^{-1}$ almost always underestimate the true values obtained by dividing the total outflowing mass by the age of the outflow. The discrepancy is typically only a factor of $<2$ but can be much higher when the outflow is young. Estimates of momentum and energy rates are similarly lower.
\end{itemize}

\begin{acknowledgements} 
We thank Roberto Maiolino and Sergei Nayakshin for their useful comments and discussion on the original draft. MAB acknowledges support from the Science and Technology Facilities Council (STFC). KZ and MT are funded by the Research Council Lithuania grant no. S-MIP-24-100. Some simulations have been performed on Galax, the computing cluster of the Centre for Physical Sciences and Technology in Vilnius, Lithuania.
\end{acknowledgements}

\bibliographystyle{mnras}
\bibliography{References}

\label{lastpage}
\end{document}